\documentclass[iop, apj, twocolappendix, numberedappendix, appendixfloats]{emulateapj}
\usepackage{threeparttable}
\usepackage{multirow}
\usepackage{times}
\usepackage{enumitem}
\usepackage{url} 
\usepackage{amsmath,bm}

\shorttitle{Photometric Re-calibration of SMSS DR2}
\shortauthors{Yang Huang et al.}
\begin{document}

\title{Milky Way Tomography with the SkyMapper Southern Survey. II. Photometric Re-calibration of SMSS DR2}

\author{Yang Huang\altaffilmark{1}}
\author{Haibo Yuan\altaffilmark{2}}
\author{Chengyuan Li\altaffilmark{3,4}}
\author{Christian Wolf\altaffilmark{5,7}}
\author{Christopher A. Onken\altaffilmark{5,7}}
\author{Timothy C. Beers\altaffilmark{6}}
\author{Luca Casagrande\altaffilmark{5,9}}
\author{Dougal Mackey\altaffilmark{5}}
\author{Gary S. Da Costa\altaffilmark{5}}
\author{Joss Bland-Hawthorn\altaffilmark{8,9}}
\author{Dennis Stello\altaffilmark{10,8,11, 9}}
\author{Thomas Nordlander\altaffilmark{5,9}}
\author{Yuan-Sen Ting\altaffilmark{12,13,14,5}}
\author{Sven Buder\altaffilmark{5,9}}
\author{Sanjib Sharma\altaffilmark{8,9}}
\author{Xiaowei Liu\altaffilmark{1}}

\altaffiltext{1}{South-Western Institute for Astronomy Research, Yunnan University, Kunming 650500, People's Republic of China; {\it yanghuang@ynu.edu.cn, {\it x.liu@ynu.edu.cn}}}
\altaffiltext{2}{Department of Astronomy, Beijing Normal University, Beijing 100875, People's Republic of China; {\it yuanhb@bnu.edu.cn}}
\altaffiltext{3}{School of Physics and Astronomy, Sun Yat-sen University, Zhuhai 519082, People's Republic of China}
\altaffiltext{4}{Department of Physics and Astronomy, Macquarie University, Sydney, NSW 2109, Australia}
\altaffiltext{5}{Research School of Astronomy and Astrophysics, Australian National University, Canberra, ACT 2611, Australia}
\altaffiltext{6}{Department of Physics and JINA Center for the Evolution of the Elements (JINA-CEE), University of Notre Dame, Notre Dame, IN 46556, USA}
\altaffiltext{7}{Centre for Gravitational Astrophysics, Research Schools of Physics, and Astronomy and Astrophysics, Australian National University}
\altaffiltext{8}{Sydney Institute for Astronomy (SIfA), School of Physics, University of Sydney, NSW 2006, Australia}
\altaffiltext{9}{Center of Excellence for Astrophysics in Three Dimensions (ASTRO-3D), Australia}
\altaffiltext{10}{School of Physics, The University of New South Wales, Barker Street, Sydney, NSW 2052, Australia}
\altaffiltext{11}{Stellar Astrophysics Centre, Department of Physics and Astronomy, Aarhus University, Ny Munkegade 120, DK-8000 Aarhus C, Denmark}
\altaffiltext{12}{Institute for Advanced Study, Princeton, NJ 08540, USA}
\altaffiltext{13}{Department of Astrophysical Sciences, Princeton University, Princeton, NJ 08540, USA}
\altaffiltext{14}{Observatories of the Carnegie Institution of Washington, 813 Santa Barbara Street, Pasadena, CA 91101, USA}

\begin{abstract}
We apply the spectroscopy-based stellar-color regression (SCR) method to perform an accurate photometric re-calibration of the second data release from the SkyMapper Southern Survey (SMSS DR2).
From comparison with a sample of over 200,000 dwarf stars with stellar atmospheric parameters taken from GALAH+ DR3 and with accurate, homogeneous photometry from $Gaia$ DR2, zero-point offsets are detected in the original photometric catalog of SMSS\,DR2, in particular for the gravity- and metallicity-sensitive $uv$ bands.
For $uv$ bands, the zero-point offsets are close to zero at very low extinction, and then steadily increase with $E (B - V)$, reaching as large as 0.174 and 0.134\,mag respectively, at $E (B - V) \sim 0.5$\,mag.
These offsets largely arise from the adopted dust term in the transformations used by SMSS DR2 to construct photometric calibrators from the ATLAS reference catalog.
For the $gr$ bands, the zero-point offsets exhibit negligible variations with SFD\,$E(B - V )$, due to their tiny coefficients on the dust term in the transformation.
Our study also reveals small, but significant, spatial variations of the zero-point offsets in all $uvgr$ bands.
External checks using Str{\"o}mgren photometry, WD loci and the SDSS Stripe 82 standard-star catalog independently confirm the zero-points found by our revised SCR method.
\end{abstract}
\keywords{catalogs -- dust, extinction -- instrumentation: photometers -- methods: data analysis -- surveys -- techniques: imaging spectroscopy}

\section{Introduction}

Our understanding of the nature of our galaxy, the Milky Way, and the large-scale structure traced by distant galaxies throughout the Universe, has been revolutionized by the advent of large, wide-field optical and near-infrared digital imaging sky surveys, such as the Sloan Digital Sky Survey (SDSS; York et al. 2000), the Two  Micron All Sky Survey (2MASS; Skrutskie et al. 2006), the Wide-field Infrared Survey Explorer (WISE; Wright et al. 2010), the Pan-STARRS1 survey (PS1; Chambers et al. 2016), the SkyMapper Southern Survey (SMSS; Wolf et al. 2018), and the $Gaia$ mission ($Gaia$ Collaboration et al. 2016, 2018).
Numerous present- and next-generation wield-field imaging surveys are expanding this wealth of information, including the Javalambre Physics of the Accelerating Universe Astrophysical Survey (J-PAS; Benitez et al. 2014), the Javalambre Photometric Local Universe Survey (J-PLUS; Cenarro et al. 2019), the Southern Photometric Local Universe Survey (S-PLUS; Mendes de Oliveira 2019), the Stellar Abundance and Galactic Evolution survey (SAGE; Zheng et al. 2018), the Large Synoptic Survey Telescope (LSST; Ivezi{\'c} et al. 2019), and the Multi-channel Photometric Survey Telescope (Mephisto; Er et al. 2020).

Accurate and homogeneous photometric calibration is the most important requirement for these surveys, in order to translate the observed signals to an absolute physical flux scale, after removal of effects from the telescope, instrument, Earth's atmosphere, and interstellar dust.  
Accurate magnitudes and colors are needed to provide robust classifications, photometric redshift estimates for a large number of galaxies to further explore the large-structure of the Universe (e.g., Padmanabhan et al. 2007; Abbott et al. 2019), and to derive the basic stellar properties (e.g., effective temperature, metallicity, age, and luminosity) for huge numbers of stars to probe Galactic structure and the star-formation, chemical-evolution, and assembly history of the MW (e.g., Juri{\'c} et al. 2008; Ivezi{\'c} et al. 2008; Casagrande et al. 2014, 2019; Da Costa et al. 2019; Huang et al. 2019; Whitten et al. 2020).  To fulfill these ambitious scientific aims, photometric calibration at a level of $\leq $1\% accuracy is the primary challenge for current and future large-scale photometric surveys.

Traditional optical photometric calibration is based on networks of standard stars with well-determined fluxes (e.g., Landolt 1992, 2009, 2013; Stetson 2000).
However, achieving $\leq 1$\% accuracy using the standard techniques for ground-based large-scale photometric surveys is difficult for several reasons, including: 1) Significant systematic errors that can be introduced when converting the photometric system of the standard stars to the system of interest (e.g., Padmanabhan et al. 2008; Finkbeiner et al. 2016), and  2) The spatial and temporal variations of Earth's atmospheric transmission and instrumental effects can be difficult to monitor by the traditional approach (e.g., Stubbs \& Tonry 2006; Yuan et al. 2015).

Thanks to the success of large-scale digital sky surveys in the past decades, in particular SDSS, several different methods have been developed to pursue the $\leq 1$\% accuracy goal for photometric calibrations across large areas of sky.
One method is the uber-calibration technique, originally developed for SDSS  (Padmanabhan et al. 2008). This method first achieves uniform internal calibrations using overlapping observed regions, then the photometric zero-points of the entire survey are scaled to a network of well-defined standard stars.
By applying this technique to the SDSS imaging data, 1\% internal accuracy has been achieved for the $griz$ bands, and about 2\% for the $u$-band (Padmanabhan et al. 2008). 
%SHOULD WE MENTION TRACTOR FROM SCHLEGEL AS WELL ?
%I found TRACTOR is a photometry code developed by Lang (am I right?). But I did not find any calibration strategy mentioned in the document.

More recently, Yuan et al. (2015; hereafter Y15) proposed a spectroscopy-based stellar-color regression (SCR) method to provide accurate color calibrations for modern imaging surveys.
As a test, this technique was applied to the SDSS Stripe\,82 multi-epoch photometric data, and achieved very high accuracies:  $\sim 5$\,mmag in $u - g$, $\sim 3$\,mmag in $g - r$, and $\sim 2$\,mmag in $r - i$ and $i - z$.
The SCR method is very powerful and straightforward for calibrating modern large-scale photometric surveys, given the fact that most of the sky is now covered by massive spectroscopic surveys, e.g., the RAVE (Steinmetz et al. 2006), SDSS/SEGUE (Yanny et al. 2009), LAMOST (Deng et al. 2012; Liu et al. 2014), and GALAH (De Silva et al. 2015) surveys.
Moreover, with the uniform (photometric calibrations at a level of 2\,mmag; Evans et al. 2018) all-sky three-band photometry ($G$, $G_{\rm BP}$, $G_{\rm RP}$) released by $Gaia$ DR2 ($Gaia$ Collaboration et al. 2018), one can extend the SCR technique to calibrate individual photometric bands, rather than just stellar colors, as we
describe below.

The SkyMapper Southern Survey (SMSS) is an on-going digital imaging survey of the entire Southern sky (Wolf et al. 2018; Onken et al. 2019).
The survey depth is expected to be 19.7-21.7 in six optical bands ($uvgriz$).
The SMSS began operation in 2014, and includes two components: the shallow survey and the main survey. The short-exposure (5 seconds in $gr$ bands and 10-40 seconds in other bands) shallow survey reaches a depth of $\sim 18$ in all six bands; the collected images and resulting catalogs were published in SMSS DR1 (Wolf et al. 2018). The second data release has published portions of the main survey, with detection limits down to $> 21$ in the $g$- and $r$-bands (Onken et al. 2019; hereafter O19). The saturation limit is set by the shallow survey, and is roughly $uv \sim$ 9 and $griz \sim $ 10.
%%%bright limit is given by the shallow survey

According to O19, the photometric zero-points in SMSS DR2 are anchored to the all-sky homogeneous $Gaia$ DR2 photometry. 
To achieve this, O19 transformed the $G_{\rm BP}$ and $G_{\rm RP}$ magnitudes of $Gaia$ DR2 to PS1 $griz$ magnitudes using the relations provided by Tonry et al. (2018).
Deviating from the description in O19, the calibration of the released DR2 catalog actually switched to using the $griz$ magnitudes from the Tonry et al. (2018) reference catalog itself.
 These PS1 $griz$ magnitudes were then further converted to SkyMapper $uvgriz$ magnitudes, using transformations based on synthetic photometry from the stellar spectral library of Pickles (1998).
The internal tests by O19 indicated that a reproducibility of 1\% in the $uv$ bands and 0.7\% in the $griz$ bands was achieved. However, the central difficulty is that the band-pass transformations used in the calibrations may introduce potentially large zero-point systematics, for two reasons. First, they include dust-correction terms in the transformations, and the adopted values of extinction may suffer large errors (especially for low Galactic latitudes). This issue is particularly serious for the $uv$ bands, since they are extrapolated from PS1 $gi$ bands, and the coefficients of the dust terms are quite large (e.g., Casagrande et al. 2019).
Secondly, the transformations do not consider metallicity effects (which are very important for the $uv$ bands); as a result, spatial patterns in the zero-points could be introduced due to stellar-population gradients across the sky (also mentioned in O19).
 
Here, we re-calibrate the photometry from SMSS DR2 using a revised SCR method, aiming to achieve uniform photometry with accuracy better than 1\%.
The paper is structured as follows.
In Section\,2, we briefly introduce an updated SCR technique using $Gaia$ DR2 Photometry.  The data employed are described in Section\,3.
In Section\,4, we perform the photometric re-calibration of SMSS DR2, and demonstrate that accuracies at the several mili-mag level can be obtained.
A discussion and brief conclusions are presented in Section\,5.

\section{SCR with $Gaia$ DR2 photometry}

The idea of the SCR method had its origins in the spectroscopic ``star pair" (hereafter, S-P) technique (Yuan \& Liu 2012; Yuan et al. 2013).
The key steps of the SCR approach, and an example of its application to SDSS Stripe 82 photometric data, are presented in Y15.
With the all-sky accurate and homogeneous photometry achieved by $Gaia$ DR2, we now can extend this technique to calibrate the photometric magnitudes of individual filters, rather than just the stellar colors.
A brief summary of the extension of this method is as follows:

\begin{enumerate}[label=\arabic*)]

\item Define a reference field with sufficient spectroscopic targets, located in a region with low extinction, and observed under good conditions. The relations between stellar intrinsic colors and atmospheric parameters (i.e., effective temperature $T_{\rm eff}$, surface gravity, log\,$g$, and metallicity, [Fe/H]) from spectroscopic observations are then determined (where multiple previous spectroscopic survey information can be used, as desired). Here the intrinsic stellar color is a combination of the photometric band, $X$, to be calibrated and one of the  $Gaia$ photometric bands ($G$, $G_{\rm BP}$, $G_{\rm RP}$), after correcting for interstellar reddening, either from the extinction map of Schlegel et al. (1998; hereafter SFD) for high Galactic latitudes, or the results estimated from the S-P technique (Yuan et al. 2013). In the $Gaia$ era, $T_{\rm eff}$ in the relation could be replaced by $(G_{\rm BP} - G_{\rm RP})_0$, given their high photometric accuracies. 

\item The $X$-band magnitudes are predicted by the relationship found in the previous step using the atmospheric parameters, the $Gaia$ photometry, and the interstellar reddening.

\item The entire set of photometric data for the survey of interest are then internally re-calibrated to the selected reference field by comparing the observed $X$-band magnitudes to the predicted ones in color, magnitude, spatial, and/or other data spaces.

\item Finally, the internal re-calibrated photometric data for a given survey can be further linked to the standard-defined photometric system (e.g., AB or Vega) by well-defined standard stars observed in the survey.

\end{enumerate}

The accuracy of this calibration technique is limited primarily by two issues: 1) the precision with which one can predict the $X$-band magnitude and, 2) the number of stars used to calibrate a certain survey field/area.
The first issue is dependent on the precision of the stellar atmospheric parameters and the particular photometric band.
Typically, the precision is about 0.02-0.03\,mag for predicting blue bands and 0.01\,mag for optical/near-infrared bands, when the metallicity precision is 0.10-0.15\,dex (can be achieved by most of the current large-scale spectroscopic surveys).
In this sense, we require at least 25 and 10 stars to calibrate blue and optical/near-infrared bands to the accuracy within 5 milli-mag (including the uncertainties of the estimates of  reddening) in a certain field of a survey.
The requirements on the aforementioned number of stars with precise atmospheric parameters (especially [Fe/H]) now is fully achieved by modern massive spectroscopic surveys (e.g., the LAMOST\footnote{LAMOST has obtained precise atmospheric parameters for over 3 million FGK dwarf stars roughly evenly distributed between $-10^{\circ}$ and $60^{\circ}$ at declination; in its  latest international data release (\url{http://dr6.lamost.org/}). 
With this dataset, one can have typically over 30-50 stars in a field of 20\arcmin$\times$20\arcmin (corresponding to a 4k$\times$4k CCD with pixel scale about 0.3\arcsec).
With such coverage, we can even examine the photometric zero-points of a survey produced in the scale of a single chip (e.g., instrumental effects like flat-fielding, pixel cross-talk, and ghosting).} survey in the northern sky and GALAH in the southern sky).
In this paper, we apply this revised SCR technique to calibrate the photometric zero-points of SMSS DR2 as an example.
As a next step, this method will be applied to more other large-scale photometric surveys (e.g., J-PLUS; Yuan et al. in preparation).

\section{Data}

In the current work, we re-calibrate the SMSS\,DR2, released in 2019 (O19).  SMSS DR2 provides photometry in at least some filters (at either the shallow- or main-survey depth), for nearly the entire Southern Hemisphere (over 21,000 deg$^{2}$), and , for the first time, data from the deep main survey in all six filters for over 7,000 deg$^2$.  In total, over 500 million unique sources and 5 billion individual detections from 120,000 images are contained in the released catalogs. As mentioned earlier, internal reproducibility tests indicated that an accuracy of 1\% in $uv$, and 0.7\% in $griz$, respectively, was achieved.

To enable our SCR method to perform photometric re-calibration of SMSS DR2, the photometric data of $Gaia$ DR2 ($Gaia$ Collaboration et al. 2018) and the stellar atmospheric parameters ($T_{\rm eff}$, log\,$g$, [Fe/H]) given by GALAH+ DR3 (Buder et al. 2020) are used in this work.
The GALAH+ DR3 catalog consists of stars observed by the GALAH,
TESS-HERMES (Sharma et al. 2018), and K2-HERMES (Sharma et al. 2019) surveys.
$Gaia$ DR2 has released three photometric bands ($G$, $G_{\rm BP}$, $G_{\rm RP}$) for over 1.3 billion stellar sources over essentially the entire sky. The typical uncertainties of $G$ and $G_{\rm BP}$/$G_{\rm RP}$ are  0.3\,mmag and 2\,mmag at $G \leq 13$, 2\,mmag and 10\,mmag at $G = 17$, and 10\,mmag and 200\,mmag at $G = 20$, respectively ($Gaia$ Collaboration et al. 2018).  The zero-points of $Gaia$ photometry are very stable across the entire sky, with a precision of a few mmag (Evans et al. 2018).

Stellar parameters ($T_{\rm eff}$, log\,$g$, [Fe/H], $v_{\rm mic}$, $v_{\rm broad}$, $v_{\rm rad}$) and over 30 elemental abundances derived from $>$ 650,000 spectra for $>$ 600,000 unique stars are included in GALAH+ DR3.  
In the current work, only the stellar atmospheric parameters of GALAH+ DR3 are used; the typical uncertainties are around 90\,K, 0.195\,dex, and 0.075\,dex for $T_{\rm eff}$, log\,$g$, and [Fe/H], respectively, for FGK-type stars with signal-to-noise ratio (SNR) greater than 20/1 per pixel.
The footprint of GALAH+ DR3 now covers a large portion (over 50\%) of the entire Southern sky, which is very helpful to re-calibrate the large-scale photometric behavior of SMSS DR2.
The typical magnitude range of the GALAH+ DR3 targets is from 8 to 14 in $G$-band, with a few fields extending to 16 at the faint end.

We note that the SFD $E (B - V)$ values used in the current work are corrected for a 14\% systematic over-estimate in the map, as found in previous works (e.g., Schlafly et al. 2010; Yuan et al. 2013).  
 
\section{Re-calibration of SMSS DR2}

In this section, we re-calibrate the current SMSS DR2 photometry (O19) using the revised SCR method described in Section\,2. If not specified otherwise, the values of the mean and standard deviation in the following analysis are obtained by the use of Maximum-Likelihood Gaussian fits.  We also tested the results using the biweight estimators (Beers et al. 1990), which provide robust estimates of central location and scale that are relatively insensitive to outliers, and found no significant differences. 

\begin{table*}
\centering
\caption{Color-Excess Ratios and Extinction Coefficients for $Gaia$ Passbands}
\begin{tabular}{ccccccc}
\hline
$\frac{E (G_{\rm BP} - K_{\rm s})}{E (B - V)}$&$\frac{E (G - K_{\rm s})}{E (B - V)}$&$\frac{E (G_{\rm RP} - K_{\rm s})}{E (B - V)}$&$\frac{E (G_{\rm BP} - G_{\rm RP})}{E (B - V)}$&$R_G$&$R _{G_{\rm BP}}$&$R_{G_{\rm RP}}$\\
\hline
$2.918 \pm 0.037$&$2.168 \pm 0.036$&$1.588 \pm 0.025$&$1.329 \pm 0.044$& $2.516 \pm 0.036$ & $3.266 \pm 0.037$&$1.936 \pm 0.025$\\
\hline
\end{tabular}
\end{table*}

\begin{table*}
\centering
\caption{Extinction Coefficients for SkyMaper Passbands}
\begin{tabular}{ccccccc}
\hline
$R_u$&$R_v$&$R_g$&$R_r$&$R_i$&$R _z$&Note\\
\hline
4.993&4.681&3.472&2.660&1.847&1.402&Wolf et al. (2018)\\
4.880&4.550&3.430&2.730&1.990&1.470&Casagrande et al. (2019)\\
5.075&4.733&3.407&2.685&2.030&1.618&Huang et al. (2019)\\
4.983&4.655&3.436&2.692&1.956&1.497&Mean value\\
\hline
\end{tabular}
\end{table*}

\subsection{Reddening Determinations and Coefficients}
Given that SMSS DR2 includes both low- and high-extinction regions, we apply the S-P method to GALAH+ DR3, together with $Gaia$ DR2 photometry, to estimate values of the color excesses, and also derive the extinction coefficients of the $Gaia$ passbands as a by-product.
To do so, we first define the control sample using the following criteria:
\begin{itemize}[leftmargin=*]

\item GALAH+ DR3 spectral {\it SNR\_c2\_iraf} greater than 20/1 and {\it flag\_sp\,$\leq$\,1}

\item $4000 \leq T_{\rm eff}/{\rm K} \leq 7000$, $0 < {\rm log}\,g < 5$, and $-1.0 \leq {\rm [Fe/H]} \leq +0.5$

\item $G$, $G_{\rm BP}$, and $G_{\rm RP}$ photometry available from $Gaia$ DR2, and with uncertainties smaller than 0.01\,mag

\item $K_{\rm s}$ photometry available from 2MASS, and with uncertainties smaller than 0.03\,mag

\item $BV$ photometry available from APASS DR9, and with uncertainties smaller than 0.035\,mag

\item Galactic latitudes $|b| \geq 30^{\circ}$ and SFD $E (B - V) \leq 0.02$

\end{itemize}
The metallicity cut here can reduce the effect of ignoring [$\alpha$/Fe] on predicting stellar colors to a few mmag.
For the target sample, the SNR cut is relaxed to 10/1 but with {\it flag\_sp\,$\leq$\,1}; the second to third criteria are the same as for the control sample.
Note that here we mainly want accurate color-excess estimates from the Gaia photometry, thus the fourth and fifth criteria are not applied to the target sample.
Here, the photometry from 2MASS and APASS DR9 are included only for empirical determinations of the reddening coefficients.
In addition, the last criterion is not used, since we want to derive the extinction values for all GALAH+ DR3 targets.
With the above cuts, 17,038 and 454,083 stars are selected for the control and target samples, respectively.

\begin{figure}
\begin{center}
\includegraphics[scale=0.2,angle=0]{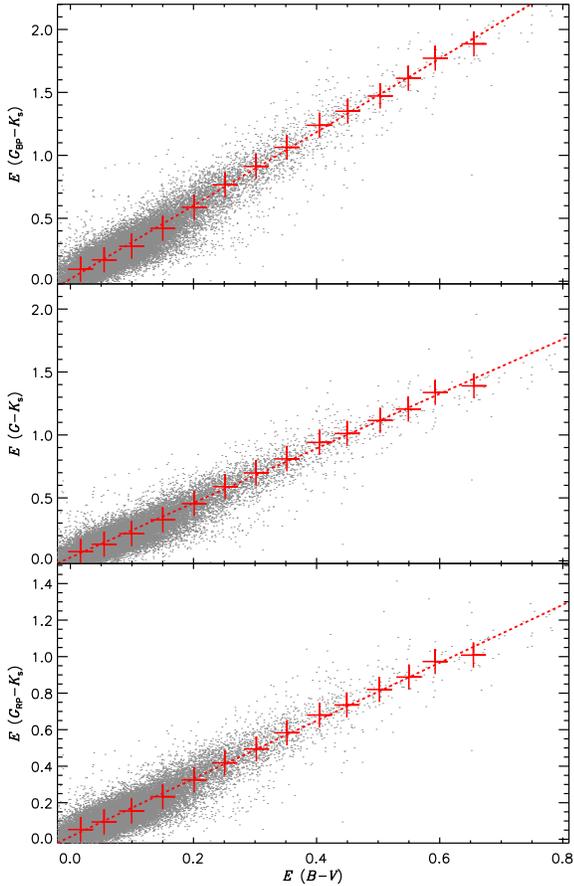}
\caption{Color excesses $E (G_{\rm BP} - K_{\rm s})$, $E (G - K_{\rm s}$), and  $E (G_{\rm RP} - K_{\rm s})$, as a function of $E (B - V)$.
Gray dots indicate data deduced from the individual stars with the same photometric qualities as the control sample and GALAH {\it flag\_sp\,$\leq 1$}.
Blue plus signs indicate mean values obtained by binning the data points into 14 groups with a bin size of 0.05\,mag in $E (B - V)$. 
The red dashed lines are first-order polynomial fits to the red plus signs, where each point carries equal weight.}
\end{center}
\end{figure}

\begin{figure*}
\begin{center}
\includegraphics[scale=0.31,angle=0]{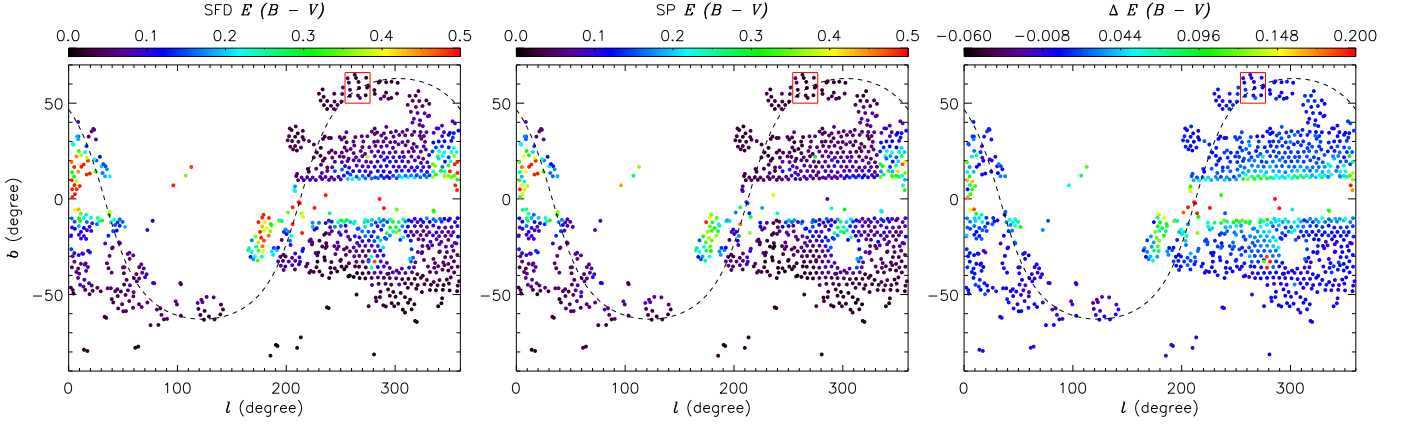}
\caption{$E (B - V)$ distribution from the SFD map (left panel) and the S-P method (middle panel; see Section\,4.1) for the (non-contiguous) GALAH+ DR3 footprint. The right panel shows the difference of $E (B - V)$ between the SFD map and the S-P estimates. 
Each dot represents a sub-field of about 3.66 deg$^2$, derived from over 400,000  GALAH+ DR3 targets (see Section\,4.1 for details) using the HEALPix algorithm (Go{\'r}ski et al. 2005). The color of each dot indicates the mean value of $E (B - V)$ (left two panels) or the $E (B - V)$ difference (right panel).  The red box marked in each panel indicates the reference field selected for our SMSS DR2 photometric re-calibration exercise (see Section\,4.2).}
\end{center}
\end{figure*}

\begin{figure}
\begin{center}
\includegraphics[scale=0.4,angle=0]{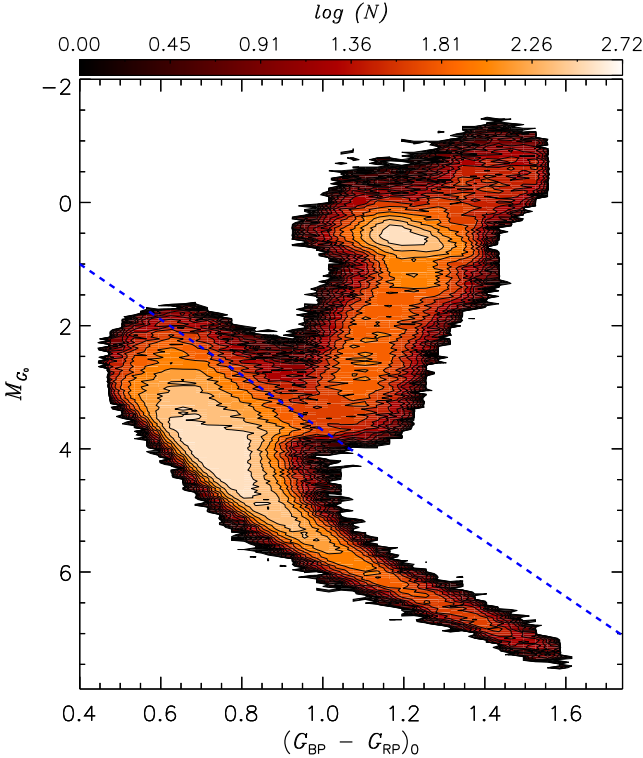}
\caption{Absolute magnitude versus de-reddened color diagram of stars with color excess $E (G_{\rm BP} - G_{\rm RP})$ estimated by the S-P method (see Section\,4.1). The logarithmic color scale represents the stellar number density.
The dashed line represents an empirical cut, i.e., $M_{\rm G_0} = -0.8 + 4.5(G_{\rm BP} -G_{\rm RP})_0$, used to separate dwarf and giant stars.}
\end{center}
\end{figure}

\begin{figure*}
\begin{center}
\includegraphics[scale=0.43,angle=0]{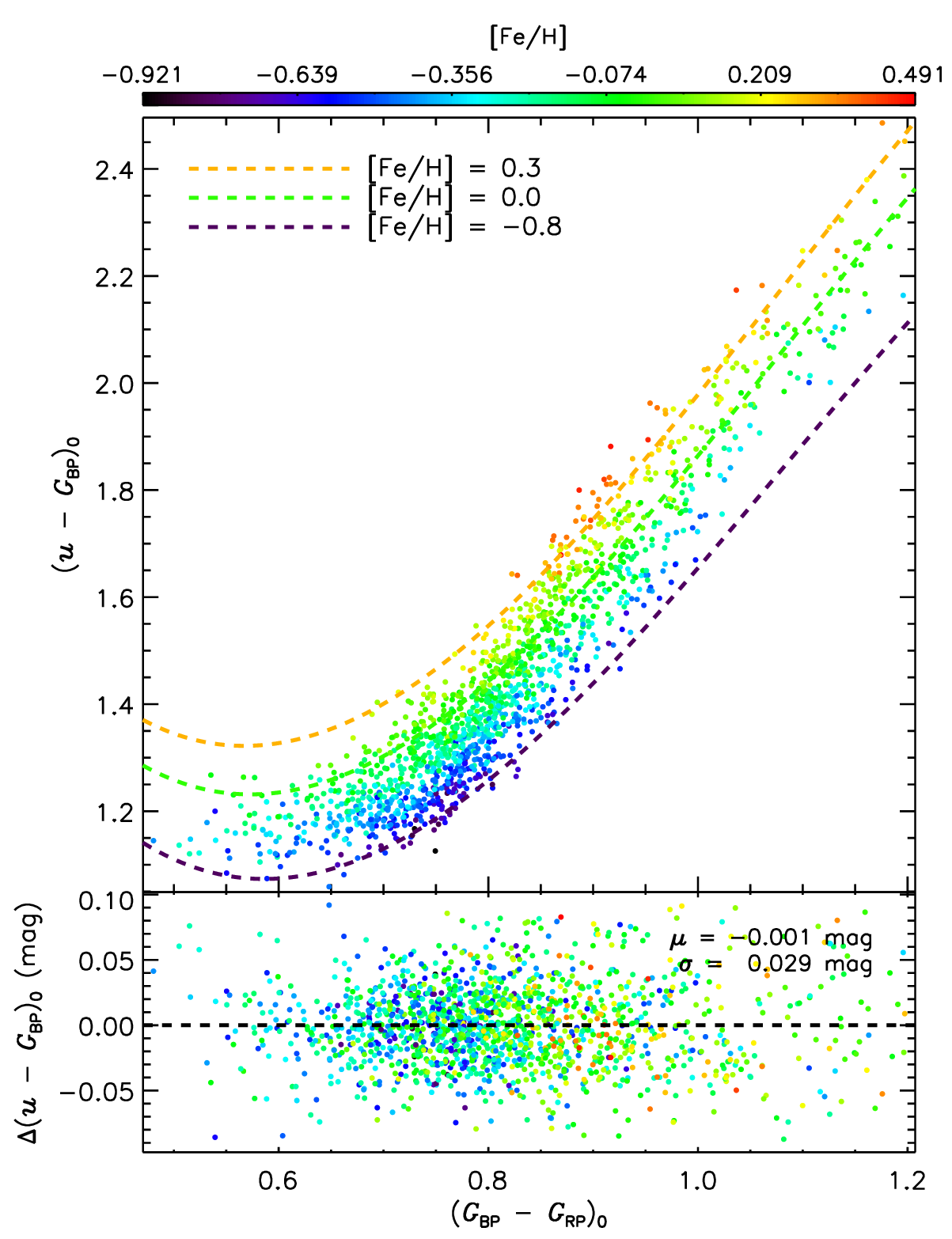}
\includegraphics[scale=0.43,angle=0]{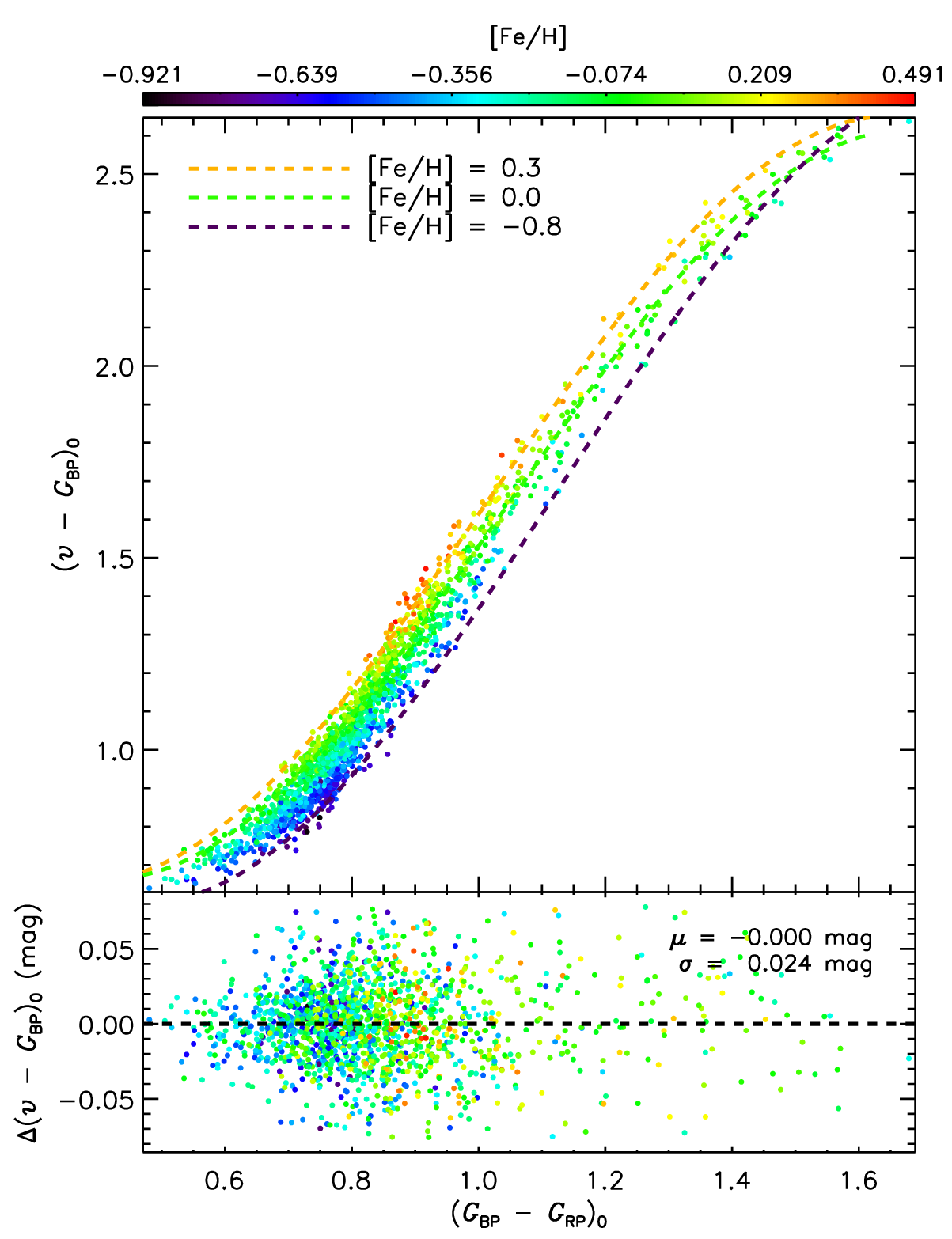}
\caption{Metallicity-dependent intrinsic color-color relations for $(u - G_{\rm BP})_0$ (left panel) and $(v - G_{\rm BP})_0$ (right panel) versus $(G_{\rm BP} - G_{\rm RP})_0$. 
The colors of the data points represent their metallicity, as indicated by the top color bar.
The dashed lines represent our best fits for selected values of [Fe/H], as marked in the top-left corner of each panel.
The bottom plot in each panel shows the best-fit residuals, with the mean and standard deviation values marked in the top-right corner.}
\end{center}
\end{figure*}

\begin{figure*}
\begin{center}
\includegraphics[scale=0.25,angle=0]{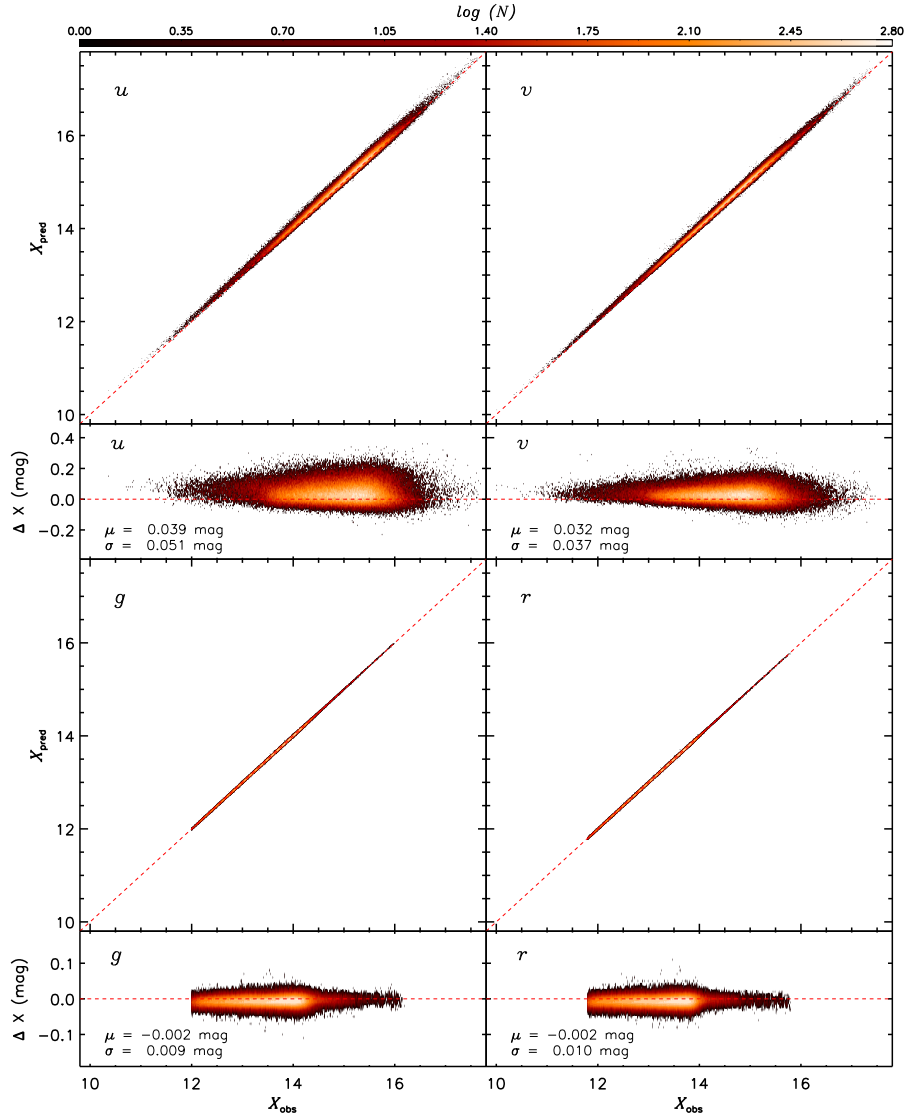}
\caption{Comparisons of the predicted SkyMapper $uvgr$ magnitudes with the observed values from SMSS DR2.
The bottom plot in each panel shows the difference between the predicted and observed magnitudes, with the mean and standard deviation marked in the bottom-left corner.
In each panel, a logarithmic red-scale map of the stellar number-density scale is shown.
Large zero-point offsets are seen for the $uv$ bands, as expected.}
\end{center}
\end{figure*}

\begin{figure*}
\begin{center}
\includegraphics[scale=0.36,angle=0]{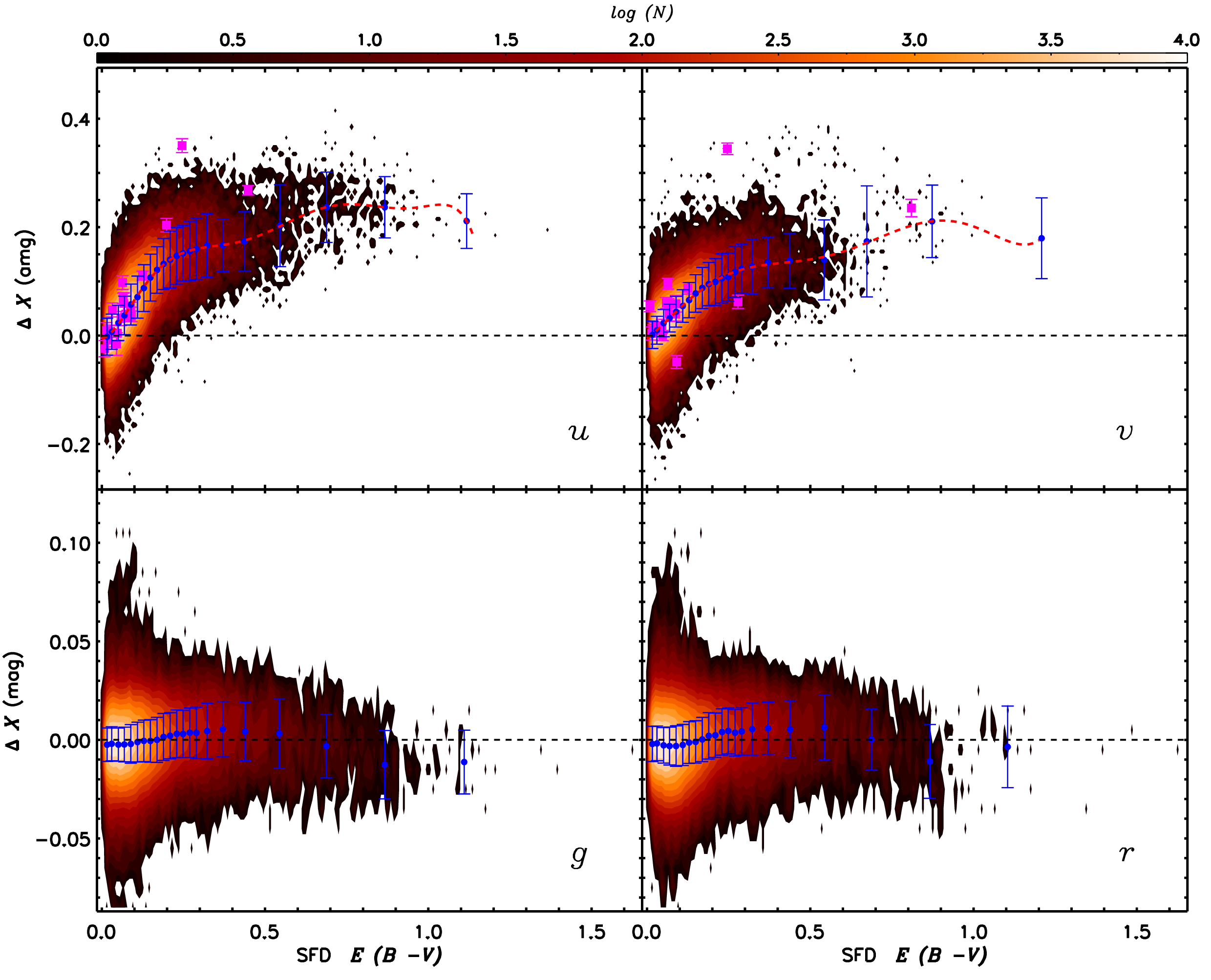}
\caption{Magnitude offsets (predicted minus observed), as a function of SFD $E (B - V)$, for $uvgr$. 
The blue dots in each panel represent the mean values of each SFD $E (B - V)$ bin; the bin ranges are given in Table\,3.
The error bars indicate the standard deviations of the magnitude differences of each SFD $E (B - V)$ bin.
The logarithmic red scale shown in each panel represents the stellar number density.
The dashed lines mark the zero value of $\Delta X$ in each panel.
The magenta squares in the top two panels represent the difference between the synthetic AB magnitudes computed for stars in the NGSL library and the SkyMapper observed values, as a function of SFD $E (B -V)$. We note an overall shift of 0.030\,mag and 0.075\,mag are adding to the difference in the $u$- and $v$-band, respectively. The error bars are given by the square root of the sum of the squares of the observed and synthetic photometric uncertainties.
The dashed red lines in the top two panels represent the best fits to the median differences as a function of SFD\,$E (B - V)$ (see Section\,4.3 for more descriptions). }
\end{center}
\end{figure*}

Here, we derive the color excesses $E (G_{\rm BP} -G_{\rm RP})$, $E (G_{\rm BP} -K_{\rm s})$ , $E (G - K_{\rm s})$, $E (G_{\rm RP} - K_{\rm s})$, and $E (B - V)$ for the target-sample stars using the S-P method  described in Yuan et al. (2013, 2015) and Huang et al. (2019).  First, the intrinsic colors of the control sample are derived from the SFD $E (B - V)$ values and the initial reddening coefficients from Chen et al. (2019) for the $Gaia$ passbands, and Fitzpartrick (1999) for the $K_{\rm s}$-band.
Secondly, we estimate the intrinsic colors of each target star by assuming that the stellar intrinsic colors vary linearly with the stellar atmospheric parameters ($T_{\rm eff}$, log\,$g$, [Fe/H]) in a small region of the parameter space\footnote{We require at least 20 stars from the control sample to derive the intrinsic colors of each target star.}, i.e., $|T_{\rm eff}^{\rm target, i} - T_{\rm eff}^{\rm control}| \leq 150$\,K, $|{\rm log} g^{\rm target, i} - {\rm log} g^{\rm control}| < 0.25$\,dex and $|{\rm [Fe/H]}^{\rm target, i} - {\rm [Fe/H]}^{\rm control}| < 0.10$\,dex.
The color excesses are then estimated by the observed colors minus the predicted intrinsic colors.
Thirdly, the resulting color excesses, $E (G_{\rm BP} - K_{\rm s})$, $E (G - K_{\rm s})$, and $E (G_{\rm RP} - K_{\rm s})$ are fitted to the color excess $E (B - V)$ by first-order polynomials, in order to obtain empirical reddening coefficients (Fig.\,1).
Finally, we iterate the above steps until the resulting reddening coefficients converge to the ones used to de-redden the control sample.

The color-excess ratios and extinction coefficients from the above procedure (adopting $A_{K_{\rm s}} = 0.348$ from Fitzpatrick 1999) for the $Gaia$ passbands are presented in Table\,1; these are in very good agreement with those derived by Chen et al. (2019). The values of the color excess $E (G_{\rm BP} - G_{\rm RP})$ for  over 400,000 GALAH+ DR3 target stars are then converted to $E (B -V)$.
By grouping the GALAH+ DR3 footprint into over 1000 fields of equal sky area (about 3.66 deg$^2$ each), we compared the mean $E (B -V)$ of stars in each field estimated by the  S-P  technique to those from SFD, in Galactic coordinates (see Fig.\,2).
The  values of $E (B -V)$ from the SFD map are in excellent agreement with those derived by the S-P  method at high Galactic latitudes and in low-extinction regions, with a typical dispersion in their difference of $\sim 0.02$\,mag.
However, one can clearly see that the SFD map over-estimates the values of $E (B -V)$ in high-extinction regions (typically at low Galactic latitudes).
The use of the SFD map in the passband transformations for photometric calibrations by SMSS DR2 will therefore contribute potentially large systematic offsets in the zero-points for high-extinction regions.
 Fig.\,3 shows the absolute magnitude versus de-reddened color diagram of the GALAH+ DR3 targets. Here, the distances estimated by Bailer-Jones et al. (2018), using the $Gaia$ DR2 parallaxes, are employed to derive the absolute magnitudes, $M_{G_{0}}$ (and requiring relative parallax errrors smaller than 30\%).

If not specified otherwise, the color excesses $E (G_{\rm BP} - G_{\rm RP})$ estimated by the S-P method are adopted in the following reddening corrections.
For the extinction coefficients, we adopt the values in Table\,1 for the $Gaia$ passbands.
For SkyMapper passbands, reddening coefficients are collected from three recent studies (Wolf et al. 2018; Casagrande et al. 2019; Huang et al. 2019; see Table\,2).
The maximum differences of the extinction coefficients amongst the three studies to the their mean values are 3.9, 1.1, 1.9, 2.6, 9.4 and 14.4 per cent for the $uvgriz$ bands, respectively.
For the reddening coefficients of the $uvgr$ bands, the mean values of the three studies are finally adopted, given their minor discrepancies amongst different studies.
For the $iz$ bands, the differences of reddening coefficients amongst the three studies are quite large.
We therefore only present photometric re-calibrations by our SCR method for SkyMapper $uvgr$ bands in the current work.
%The re-calibration results by the SCR method for $iz$ bands are however not conclusive if considering such a large extinction coefficient difference (10-15 per cent) amongst different studies.
The re-calibration results by the SCR method for the $iz$ bands are consistent with no trends just like for the $gr$ bands if we adopt the mean value of the reddening coefficients from Table 2.
 However, because of the large range in extinction coefficients (10-15 per cent) among different studies, this is no proof of absence of a zero-point trend with reddening. 
Adopting the smallest or largest values for $R_i$ and $R_z$ would allow for trends of up to $-0.031$/$+0.032$ and  $-0.025$/$+0.048$\,mag at $E (B - V) \sim 0.50$, respectively.

\subsection{Predictions of SkyMapper $uvgr$ Magnitudes}

In this section, the SkyMapper magnitudes in the $uvgriz$ bands are predicted using the SCR technique.  We only use dwarf stars, thus the effects of surface gravity log\,$g$ ($\sim 4.0$) on the stellar intrinsic colors can be neglected.
The dwarf stars are selected using an empirical cut in the $(G_{\rm BP} - G_{\rm RP})_0$\,--\,$M_{G_0}$ plane (see Fig.\,2).
In total, 234,261 dwarf stars with $E (G_{\rm BP} - G_{\rm RP})$ estimates, $Gaia$ DR2 photometry, SMSS photometry, and GALAH+ DR3 atmospheric parameters are selected.

As mentioned in Section\,3, we need to define a reference field with sufficient GALAH+ DR3 targets. In addition, this field should be located in a region of low extinction, so that the SMSS DR2 photometric calibrations are not affected by the dust terms. The region of $254^{\circ} \leq l \leq 277^{\circ}$ and $50^{\circ} \leq b \leq 66^{\circ}$ is defined as the reference field for this exercise.
The dwarf stars in this reference field satisfying the following criteria are then used to build the metallicity-dependent intrinsic color-color relations:

\begin{itemize}[leftmargin=*]
\item SFD $E (B - V) \leq 0.03$

\item Good photometric quality from SMSS DR2: {\it x\_ngood $\ge 1$, {\it x\_flags $\le 3$}}, {\it e\_x\_psf $\le 0.05$}\,mag, {\it class\_star $\ge 0.9$}, here $x$ represents the individual $u$/$v$/$g$/$r$ bands

\item $g \geq 12$, $r \geq 11.8$, $i \geq 11.0$ and $z \geq 11.0$, in order to avoid saturation

\item Good photometric quality from $Gaia$ DR2: The photometric uncertainties in $G$, $G_{\rm BP}$ and $G_{\rm RP}$ bands are all required to be smaller than 0.01\,mag

\end{itemize}

In total, over 1500 GALAH dwarf stars are selected in the reference field.
Using these dwarf stars, we have performed third-order 2-D polynomial fitting (with 10 free parameters) to their de-reddened $u - G_{\rm BP}$, $v - G_{\rm BP}$, $g - G_{\rm BP}$ and $r - G_{\rm RP}$ colors, as a function of $(G_{\rm BP} - G_{\rm RP})_0$ and [Fe/H].
The scatter of the fit residuals are 0.029, 0.024, 0.008 and 0.008\,mag for the colors $u - G_{\rm BP}$, $v - G_{\rm BP}$, $g - G_{\rm BP}$ and $r - G_{\rm RP}$, respectively.
As an example, Fig.\,4 shows the fits and residuals for the colors $u - G_{\rm BP}$ and $v - G_{\rm BP}$.

We apply the above metallicity-dependent intrinsic color-color relations to all the dwarfs selected above to predict SkyMapper $uvgr$ magnitudes, using $Gaia$ photometry, GALAH [Fe/H], and the $E (G_{\rm BP} - G_{\rm RP})$ color excesses.
Here, we require those dwarfs to satisfy the second to fourth criteria listed above for stars in the reference field.
In total, SkyMapper $uvgr$ magnitudes for over 200,000 dwarf stars are predicted in this manner.
Fig.\,5 compares the predicted $uvgr$ SkyMapper magnitudes to the observed magnitudes from SMSS DR2.
From inspection of this figure, significant zero-point systematics are detected for the $uv$ bands when comparing the predicted and observed magnitudes.
For the $gr$ bands, the predicted magnitudes agree with the observed magnitudes to within 10\,mmag, in excellent agreement with the accuracy reported by O19.

\begin{figure*}
\begin{center}
\includegraphics[scale=0.38,angle=0]{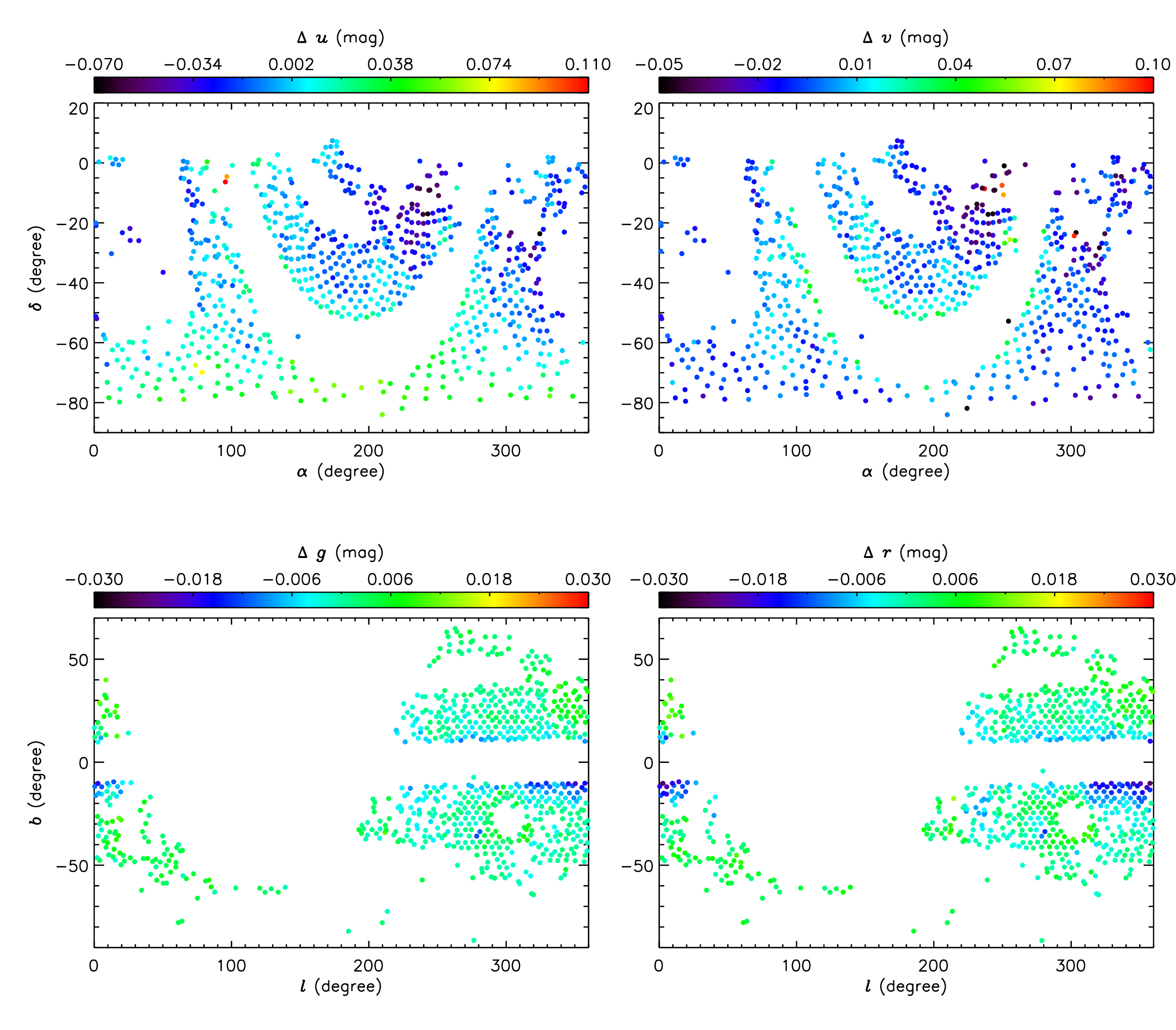}
\caption{Distributions of the mean magnitude differences (predicted minus observed, corrected for the reddening-dependent systematics), shown in the equatorial coordinate system for the $uv$ bands (the top two panels), and in the Galactic coordinate system for the $gr$ bands (the bottom two panels).
Each dot represents a field of about 3.66 deg$^2$, using the HEALPix algorithm (Go{\'r}ski et al. 2005). 
The color of each dot shows the mean value of the magnitude differences for the specified bands.}
\end{center}
\end{figure*}

\begin{figure*}
\begin{center}
\includegraphics[scale=0.38,angle=0]{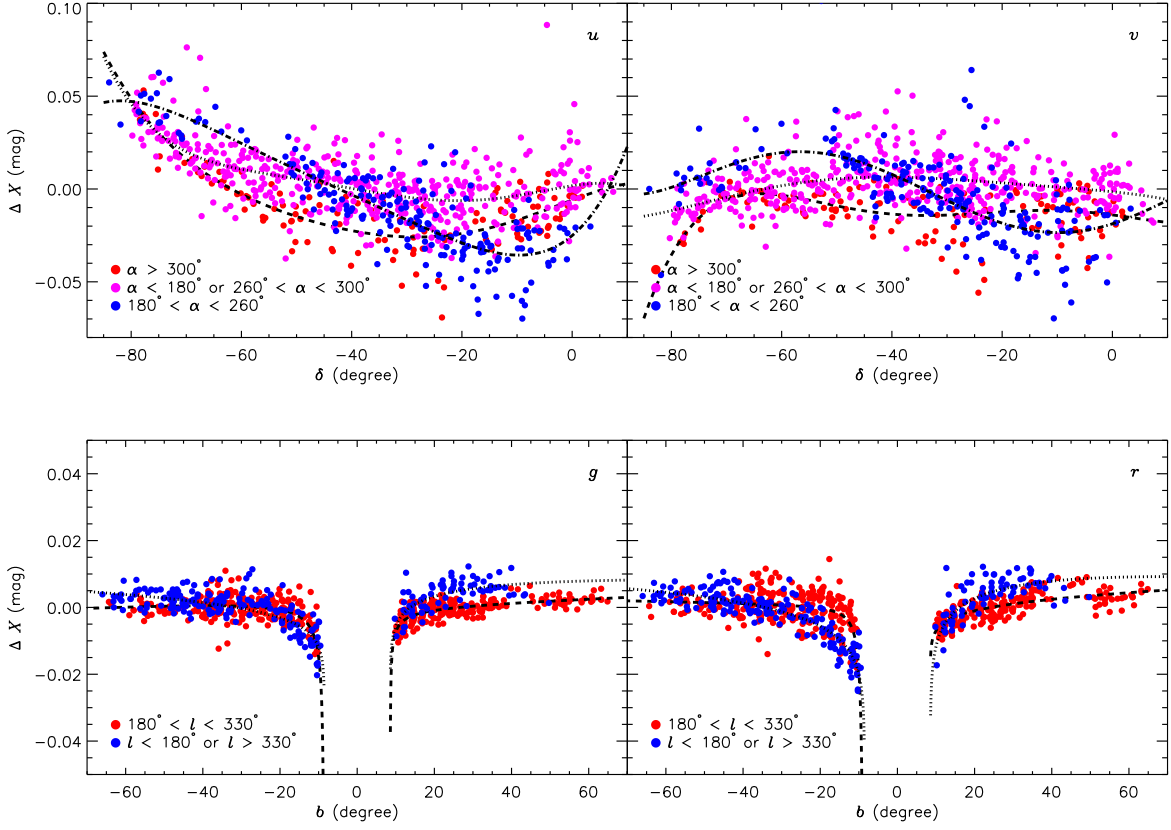}
\caption{The mean  magnitude differences (same as those shown in Fig.\,7), as a function of Declination ($\delta$), for different Right Ascension ($\alpha$) bins ($uv$ bands, the top two panels) or Galactic latitude ($b$), for different Galactic longitude ($l$) bins ($gr$ bands, the bottom two panels).
The different colors indicate data points for different $\alpha$ or $l$ ranges.
The different lines represent the best fits described in Section\,4.4.}
\end{center}
\end{figure*}

\begin{figure*}
\begin{center}
\includegraphics[scale=0.38,angle=0]{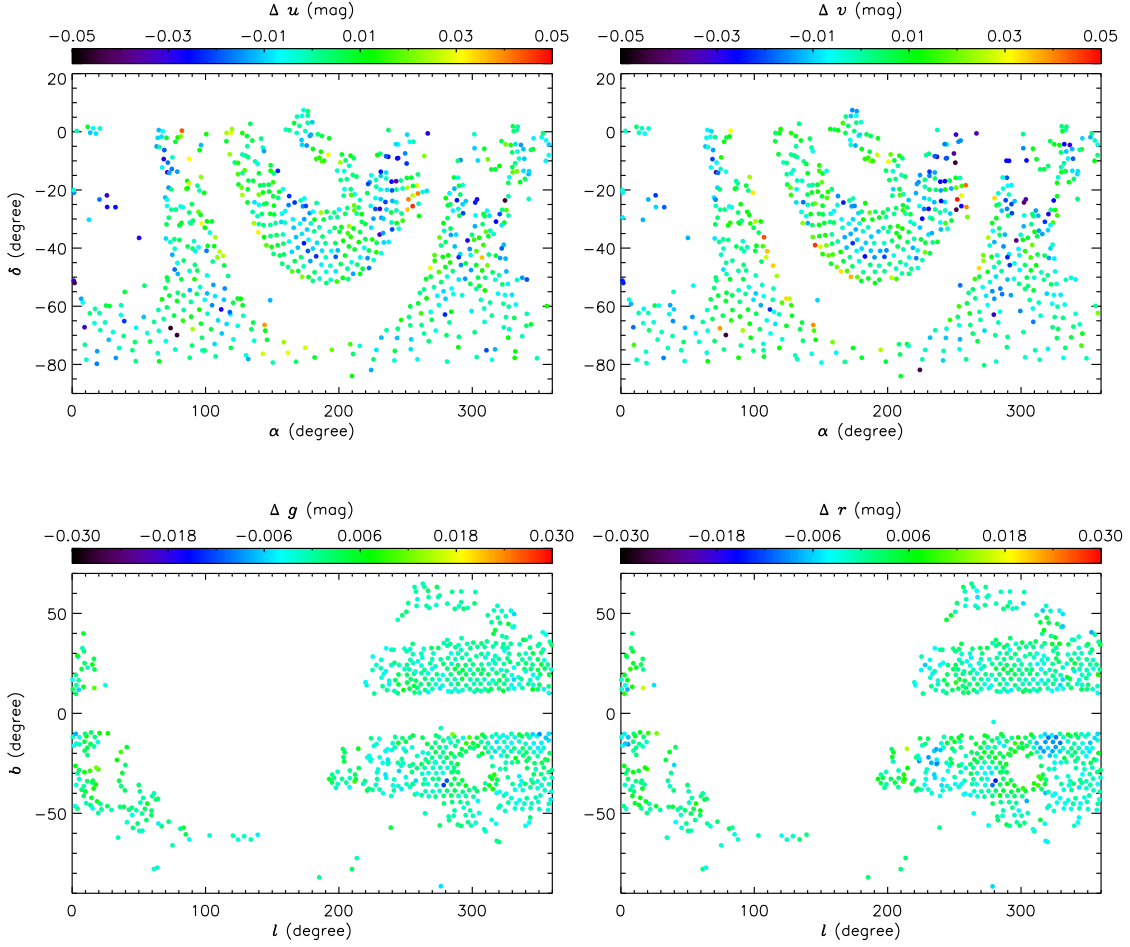}
\caption{Same as Fig.\,7, but the observed magnitudes have been corrected for both reddening-dependent and spatial systematics.}
\end{center}
\end{figure*}

\begin{figure*}
\begin{center}
\includegraphics[scale=0.38,angle=0]{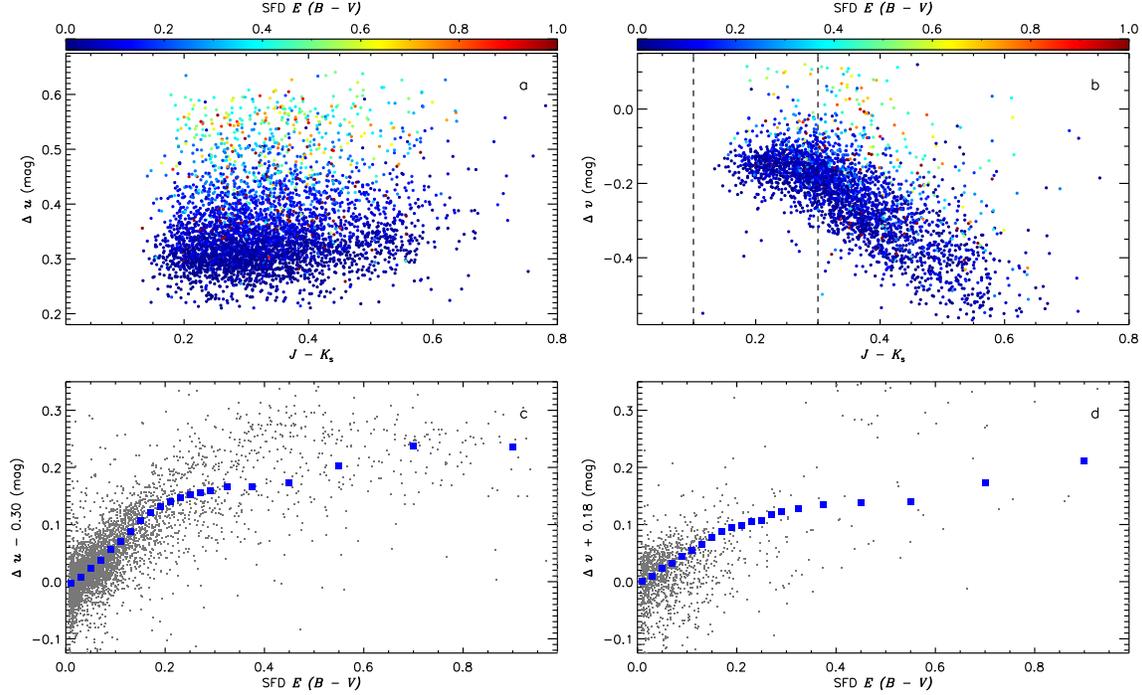}
\caption{Comparison between Str{\"o}mgren and SMSS $uv$ bands. The top two panels show the magnitude differences (Str{\"o}mgren minus SMSS), as a function of stellar colors $J - K_{\rm s}$ for the $u$ (panel a) and $v$ (panel b) bands.
The color of each dot indicates the value of SFD $E (B - V)$, represented by the top color bar.
The bottom two panels present the magnitude differences (Str{\"o}mgren minus SMSS), as a function of SFD $E (B - V)$. Here we note for the $v$-band, only stars with $0.1 \leq J - K_{\rm s} \le 0.3$ (see the dashed lines marked in panel b) are shown in panel d. 
Stars in this narrow color range exhibit a nearly constant magnitude difference in the $v$-band. The over-plotted blue squares are the zero-point offsets detected by our SCR technique (see Table\,3 and Fig.\,6).}
\end{center}
\end{figure*}

\begin{figure*}
\begin{center}
\includegraphics[scale=0.40,angle=0]{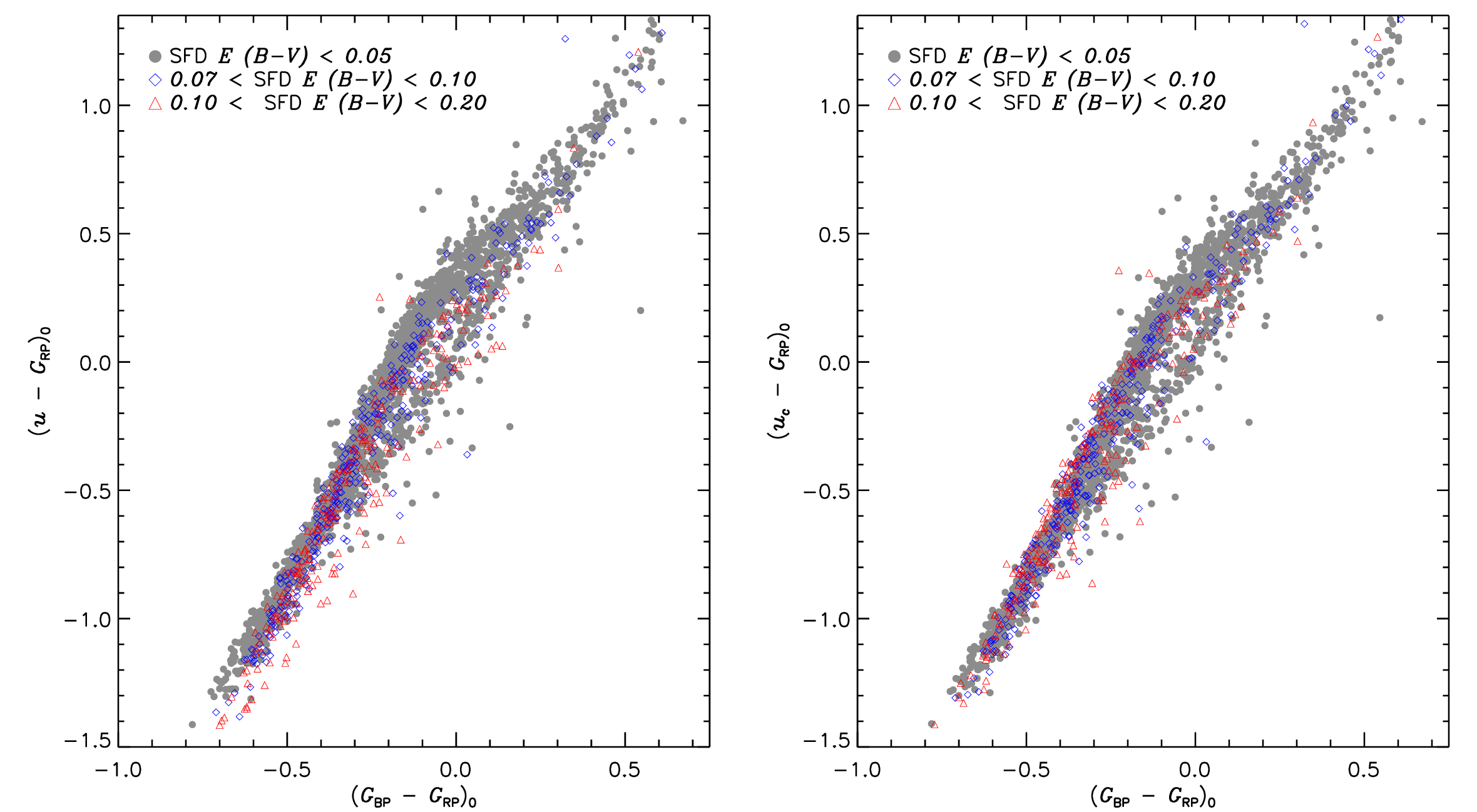}
\includegraphics[scale=0.40,angle=0]{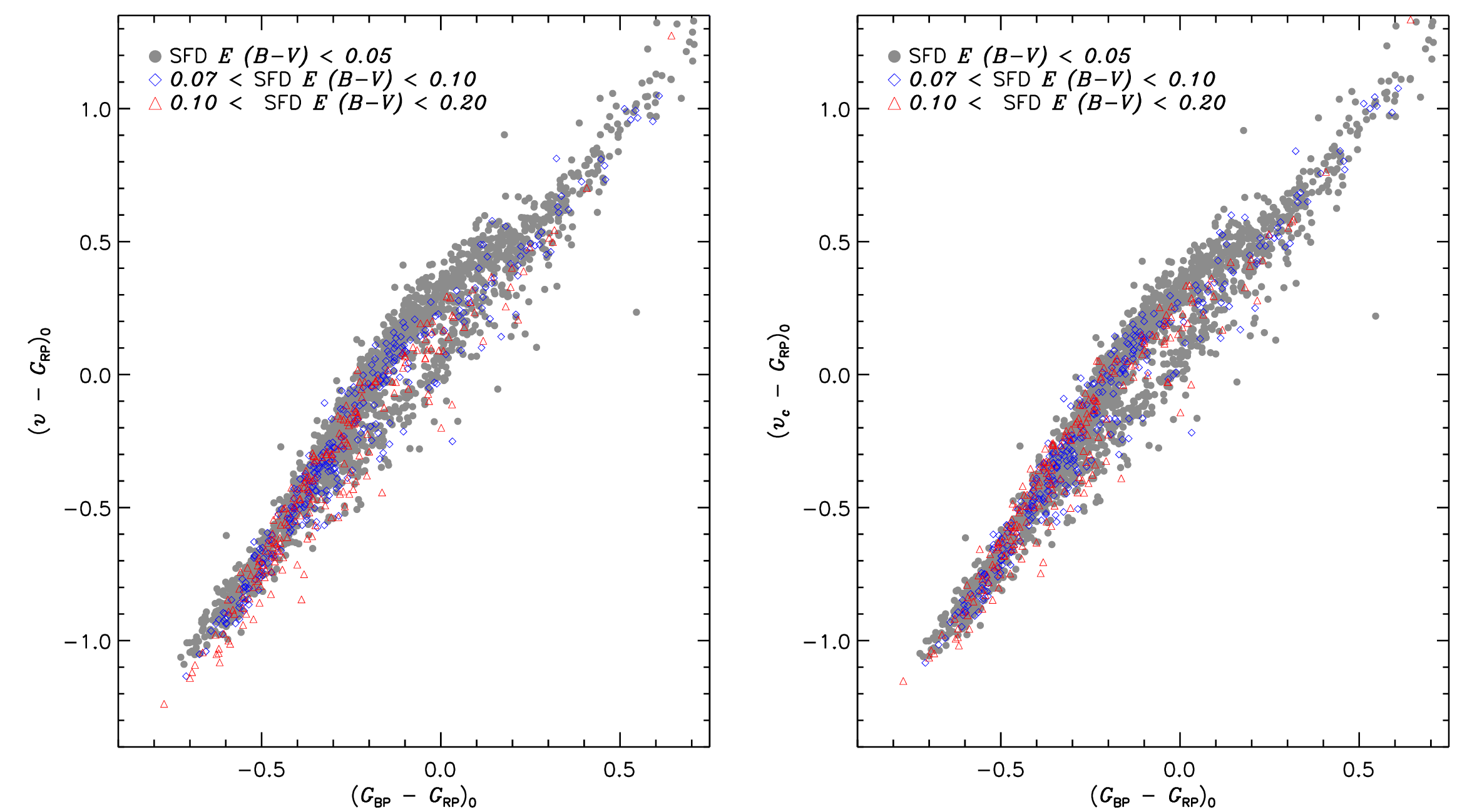}
\caption{The WD loci $(u/v - G_{\rm RP})_0$ versus $(G_{\rm BP} - G_{\rm RP})_0$. The top two panels show the locus for the $u$-band, with the left one from the original SMSS DR2 photometry, and the right one from the re-calibrated photometry. The bottom two panels are the same as in the top panel, but for the $v$-band. The gray dots, blue diamonds, and red triangles represent WDs with SFD $E (B - V)$ of $< 0.05$, between 0.07 and 0.10, and between 0.10 and 0.20, respectively.}
\end{center}
\end{figure*}

\begin{figure*}
\begin{center}
\includegraphics[scale=0.40,angle=0]{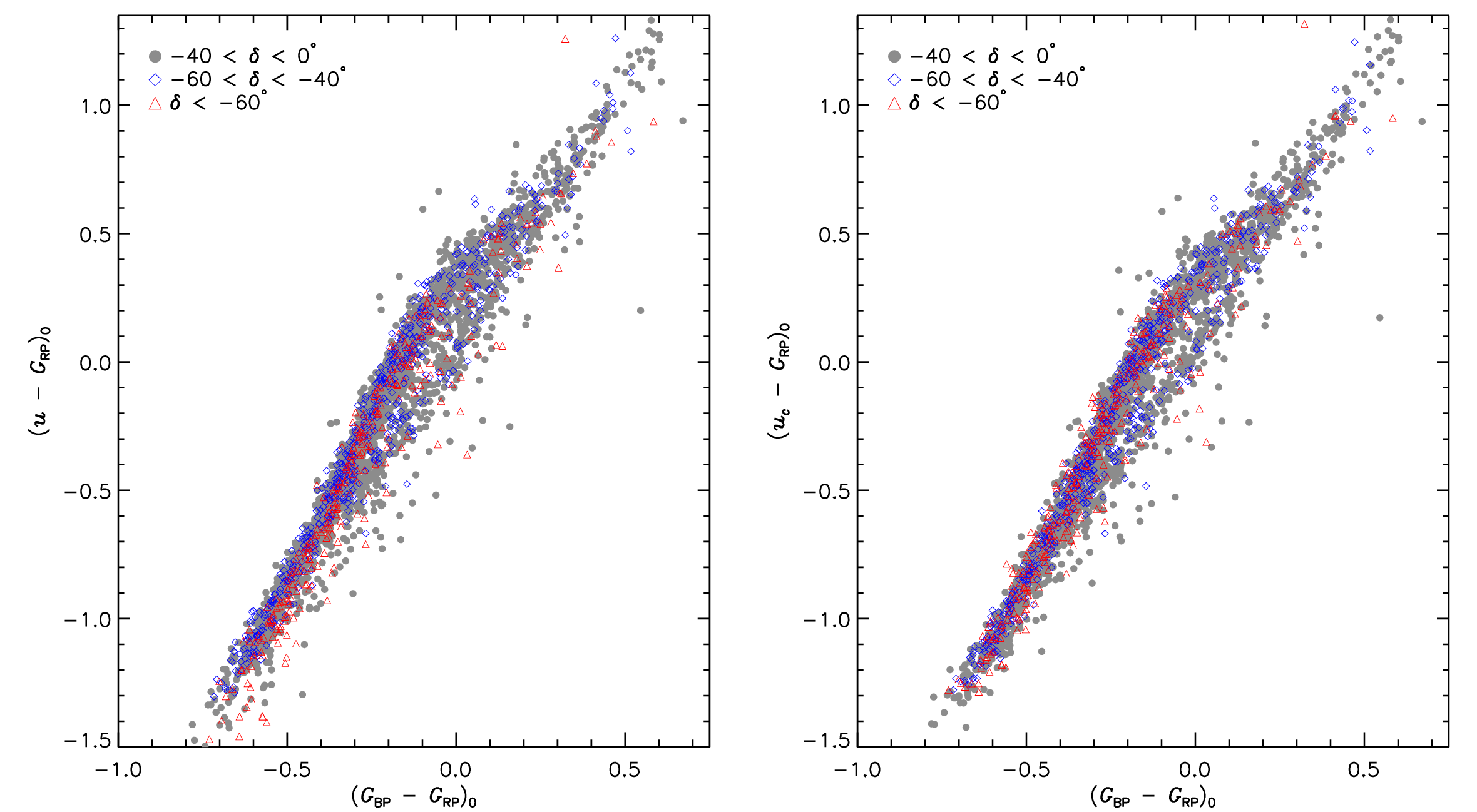}
\caption{The WD loci $(u - G_{\rm RP})_0$ versus $(G_{\rm BP} - G_{\rm RP})_0$ for different Declination ($\delta$) bins, with the left panel from the original SMSS DR2 photometry and the right panel from the re-calibrated photometry. 
Gray dots, blue diamonds, and red triangles represent WDs located at $-40 < \delta < 0^{\circ}$,  $-60 < \delta < -40^{\circ}$ and $\delta < -60^{\circ}$, respectively.}
\end{center}
\end{figure*}

\subsection{Dependence on Reddening}
As mentioned previously, the zero-points of the SkyMapper photometry are expected to  exhibit dependencies on the extinction values, due to the dust terms included in the passband transformations used for constructing photometric calibrators from $Gaia$ DR2.
Thus, we first check the differences between the predicted $uvgr$ magnitudes and the observed magnitudes, as a function of SFD\,$E (B - V)$; the results are listed in Table\,3 and shown in Fig.\,6.
%As expected, the mean differences are nearly zero for SFD $E (B - V) \leq 0.04$, and then gradually increase/decrease with increasing SFD $E (B - V)$ in the $uvgr$ bands.
For $uv$ bands, the mean differences are nearly zero for SFD $E (B - V) \leq 0.04$, and then gradually increase with increasing SFD $E (B - V)$.
These systematics can reach values as large as 0.174 and 0.134\,mag for $uv$, respectively, at SFD $E (B -V) \sim 0.50$.
Such large reddening-dependent zero-point systematics are actually expected, since $uv$ magnitudes of the calibrators in O19 are extrapolated from the redder optical bands, and the coefficients of the dust terms in the passband transformations are quite large (ten to a hundred times larger than those for the redder bands).
For larger SFD $E (B - V)$ ranges ($> 0.3$, the zero-point systematics tend to be stable or slightly declining.
This is because O19 adopted extinction values from the SFD map for $E (B - V) < 0.3$, and from the combination of the SFD map and the $Gaia$ estimates of $A_{G}$ (Andrae et al. 2018) for  $E (B - V) \ge 0.3$ in the passband transformations.
The partial use of  $A_{G}$ estimates from $Gaia$ for high-extinction regions could reduce the over-estimates of the dust terms from the SFD map, thus the zero-point systematics tend to be flat/declining at higher $E (B - V)$.
However, it should be kept in mind that the uncertainties in the $Gaia$ $A_G$ estimates can be quite large, resulting in  large scatter of the magnitude differences in high-extinction regions.
The zero-points of $gr$-bands are largely within 5\,mmag, and exhibit minor variations with SFD $E (B - V)$, due to their negligible coefficients on the dust term in the transformation.

For ease of correction, we also provide seventh-order polynomial fittings to the $u$-band and $v$-band  zero point trends as a function of SFD $E (B - V)$ (i.e., reported values in Table\,3), respectively. 
The best-fits could describe the reddening dependent zero-point trends very well for $E (B - V) < 1.20$, leaving a 2 milli-mag standard deviation (see Fig.\,6).

%Again, as expected, the reddening-dependent zero-point systematics are very large for the $uv$ bands, since they are extrapolated from the redder optical bands, and the coefficients of the dust terms in the passband transformations are quite large (ten to a hundred times larger than those for the redder bands).
%Unlike the other five bands, the $g$-band zero-point shows a (very weak) negative trend with $E (B - V)$.
%This is easily understood, as the coefficient on the dust term of the $g$-band is positive, while the coefficients of the other five bands are all negative (see O19).
%The zero-point of the $r$-band is largely within 6\,mmag, and exhibits minor variations with SFD $E (B - V)$, due to its negligible coefficient on the dust term in the transformation. 
%For this reason, we believe that the reddening-dependent zero-point systematics of the SMSS DR2 photometry largely arise from the dust terms in the passband transformations used for the photometric calibration by O19. 
%Note that the correction of the reddening-dependent zero-point from Table\,2 for the SMSS DR2 is actually made under the assumption that the over-estimated value of the extinction by the SFD map is a function of SFD $E (B - V)$, and any variations with other factors (e.g., positions) are marginalized over.

As an independent check, we cross-match the SMSS DR2 to the STIS Next Generation Spectral Library (NGSL)\footnote{\url{https://archive.stsci.edu/prepds/stisngsl/}}. 
A total of 39 stars in common are found within a 6\arcsec\ match radius, and most of those stars are saturated in the $griz$ bands.
We thus only integrate the synthetic AB magnitudes of those stars in common for the $uv$ bands, using the spectra from the NGSL.
After removing variable stars included in the General Catalog of Variable Stars\footnote{\url{http://www.sai.msu.su/gcvs/gcvs/}}, excluding red stars $(G_{\rm BP} - G_{\rm RP})_0 > 1.1$, stars with potential stellar flares/activity or/and long-term variability, and requiring good photometric quality (photometric errors smaller than 0.02\,mag, no source within 15\arcsec, and without bad SkyMapper flags), 21 and 19 stars are left in the $u$ and $v$ bands, respectively.
The uncertainties in the synthetic magnitudes are all around 0.01\,mag, and none are larger than 0.02\,mag. 
As shown in Fig.\,6, the differences between the synthetic AB magnitudes the observed SkyMapper stars in the $uv$ bands exhibit variations with SFD $E (B -V)$ that are very similar to the trends found using the SCR technique.

Finally, we point out that the Galactic-latitude dependent zero-point offsets of the $uv$ bands in SMSS DR1.1 found by Casagrande et al. (2019), using stellar effective temperatures, are actually similar to the reddening-dependent zero-point systematics detected in the current work, since the extinction value naturally increases with decreasing Galactic latitude.
The fit coefficients are presented in Table\,4.

\subsection{Spatial Variations}

After correcting for the reddening-dependent zero-point systematics using the values listed in Table\,3, we further check on the spatial variations of the zero-points in SMSS DR2. To accomplish this, we divide the above $>$ 200,000 dwarfs into over 700 fields of equal sky area (about 3.66 deg$^2$), requiring at least 50 stars be present in each field\footnote{This number of stars allows us to perform photometric calibration at the few mmag level for individual fields, given the predicted accuracies of magnitudes for individual stars (see Fig.\,4 and Section\,4.2).}.
The mean values of the magnitude differences (predicted minus observed) of each field are calculated and shown in Fig.\,7.
From inspection, the zero-points of all $uvgr$ bands exhibit significant spatial variations. For the $uv$ bands, The zero-points mainly vary with Declination ($\delta$), but also show different behaviors for different Right Ascension ($\alpha$) bins.
The zero-point spatial patterns with $\delta$ of the $uv$ bands possibly arise 
 from residuals in the corrections applied for terrestrial atmospheric extinction (especially in the $u$-band).  Clear positive zero-points are seen around the South Celestial Pole ($\delta < -70^{\circ}$); this may result from large values of airmass and atmospheric extinction in the $u$-band, in particular.
 To remove the spatial patterns of the zero-points, we have performed fifth- and fourth-order polynomial fitting to the $u$-band and $v$-band zero-point systematics, as a function of $\delta$, for different $\alpha$ ranges, respectively:
 \begin{equation}
\Delta u = a_0 + a_1\delta + a_2\delta^2+a_3\delta^3+a_4\delta^4+a_5\delta^5,
 \end{equation}
  \begin{equation}
\Delta v= a_0 + a_1\delta + a_2\delta^2+a_3\delta^3+a_4\delta^4.
 \end{equation}
The fitting results and coefficients are presented in Fig.\,8 and Table\,5, respectively.

For the $gr$ bands, the zero-points are close to zero in high Galactic latitude ($b$) regions, and rapidly decrease with decreasing $b$ in low-latitude regions.
To correct this $b$-dependent systematic, we fit the zero-points, as a function of $b$, for the $gr$ bands with equations of the form: 
\begin{equation}
\Delta X = a_0 + a_1(|b| - 8.5)^{a_2} + a_3b,
\end{equation}
\noindent where $X$ represents the band under consideration (i.e., $gr$).
The fitting results and coefficients are again presented in Fig.\,8 and Table\,5, respectively.

We also checked the zero-points of SMSS DR2 photometry in other spaces, e.g., magnitudes and colors, and no significant variations were detected.

\begin{table}
\centering
\caption{Magnitude Zero-Point Offsets as a Function of SFD $E (B - V)$}
\begin{tabular}{ccccc}
\hline
SFD $E (B -V)$&$\Delta u$&$\Delta v$&$\Delta g$&$\Delta r$\\
&(mag)&(mag)&(mag)&(mag)\\
\hline
$[0.000, 0.020]$&$-0.0020$&$+0.0014$&$-0.0025$&$-0.0021$\\
$[0.020, 0.040]$&$+0.0072$&$+0.0098$&$-0.0020$&$-0.0018$\\
$[0.040, 0.060]$&$+0.0233$&$+0.0230$&$-0.0024$&$-0.0028$\\
$[0.060, 0.080]$&$+0.0371$&$+0.0322$&$-0.0024$&$-0.0031$\\
$[0.080, 0.100]$&$+0.0573$&$+0.0440$&$-0.0021$&$-0.0031$\\
$[0.100, 0.120]$&$+0.0709$&$+0.0545$&$-0.0011$&$-0.0026$\\
$[0.120, 0.140]$&$+0.0875$&$+0.0653$&$-0.0005$&$-0.0014$\\
$[0.140, 0.160]$&$+0.1069$&$+0.0777$&$-0.0006$&$-0.0010$\\
$[0.160, 0.180]$&$+0.1210$&$+0.0877$&$-0.0000$&$+0.0002$\\
$[0.180, 0.200]$&$+0.1320$&$+0.0949$&$+0.0015$&$+0.0020$\\
$[0.200, 0.220]$&$+0.1400$&$+0.0989$&$+0.0021$&$+0.0023$\\
$[0.220, 0.240]$&$+0.1469$&$+0.1055$&$+0.0030$&$+0.0040$\\
$[0.240, 0.260]$&$+0.1517$&$+0.1078$&$+0.0030$&$+0.0045$\\
$[0.260, 0.280]$&$+0.1556$&$+0.1172$&$+0.0036$&$+0.0036$\\
$[0.280, 0.300]$&$+0.1596$&$+0.1234$&$+0.0035$&$+0.0043$\\
$[0.300, 0.350]$&$+0.1662$&$+0.1271$&$+0.0044$&$+0.0054$\\
$[0.350, 0.400]$&$+0.1667$&$+0.1347$&$+0.0054$&$+0.0058$\\
$[0.400, 0.500]$&$+0.1738$&$+0.1378$&$+0.0042$&$+0.0053$\\
$[0.500, 0.600]$&$+0.2027$&$+0.1396$&$+0.0032$&$+0.0063$\\
$[0.600, 0.800]$&$+0.2367$&$+0.1737$&$-0.0031$&$+0.0001$\\
$[0.800, 1.000]$&$+0.2367$&$+0.2106$&$-0.0126$&$-0.0109$\\
$[1.000, 1.500]$&$+0.2112$&$+0.1795$&$-0.0113$&$-0.0035$\\
$[1.500, 2.500]$&$+0.1583$&$+0.1137$&$-0.0084$&$-0.0107$\\
\hline
\end{tabular}
\tablecomments {The offsets should be added to the official SMSS DR2 values.}
\end{table}

\subsection{Final Accuracies}

After properly correcting for the reddening-dependent and spatial systematics, the final mean magnitude differences for the over 700 comparison fields are shown in Fig.\,9.
From inspection, the zero-points for most of the re-calibrated regions are within 10\,mmag.  More quantitatively, the zero-points for more than 50\% of the fields are within 7.1, 6.3, 1.8 and 2.2\,mmag for $uvgr$, respectively.
For 90\% of the fields, the zero-points are better than 17.1, 16.2, 4.5 and 6.0\,mmag for $uvgr$, respectively.
As we discuss below, the accuracies of the zero-points for the SMSS DR2 photometry could be even further improved (see more discussions in Section\,5.1). 

\subsection{External Checks by Comparison with Str{\"o}mgren Photometry}
All of the above photometric systematics found for SMSS DR2 relied on the SCR method.
Similar to Casagrande et al. (2019), we can provide an independent check of our re-calibration by comparing the Str{\"o}mgren photometry with the re-calibrated SMSS photometry .

To accomplish this, the SMSS DR2 catalog is cross-matched to the Geneva-Copenhagen Survey (Olsen 1983, 1984; Holmberg et al. 2009);  over 5000 stars in common are found.
The magnitude differences, as a function of $J-K{\rm s}$, between the Str{\"o}mgren and SMSS $uv$ bands are shown in panels a and b of Fig.\,10, respectively.
Significant zero-point offsets are found, since the Str{\"o}mgren photometry is not anchored to the AB system.
For the $u$-band, the offset is almost a constant as a function of stellar colors, since the SkyMapper and Str{\"o}mgren $u$-band transmission curves are almost identical.
For the $v$-band, a significant trend is detected as a function of the stellar colors, because the Str{\"o}mgren $v$-band is shifted  $\sim$200\,{\AA} towards the red compared to the SkyMapper $v$-band.
From inspection, outliers from the sequences are mainly the stars with large SFD $E (B - V)$ values, indicating a reddening-dependent trend of the magnitude differences.
To better show such a trend, we present the magnitude differences of the $uv$ bands, as a function of SFD\,$E (B - V)$, in panels c and d of Fig.\,10, respectively, similar to the plots in Fig.\,6.
Indeed, significant reddening-dependent differences are found in both the $u$- and $v$-bands.
Moreover, the trends are in excellent agreement with the results found by our SCR technique.
In other words, our re-calibrated $uv$ photometry matches the Str{\"o}mgren bands very well, except for arbitrary shifts.

\subsection{External Checks by Comparison with White Dwarf Loci}
Here, we provide an independent check of the re-calibration by the SCR, using a white dwarf (WD) locus.
This locus is expected to be very stable and uniform for WDs at different spatial locations, and with different values of reddening. 

We first cross-match the WD catalog constructed by Gentile Fusillo et al. (2019) from $Gaia$ DR2 with the SMSS DR2 catalog.
We require the WDs to have Galactic latitude $|b| > 20^{\circ}$, allowing one to correct the reddening using the SFD map.
In addition, the WD probability ($P_{\rm WD}$ in the Gentile Fusillo catalog) is required to be higher than 75\%, and the uncertainties of $Gaia$ $G_{\rm BP}$ and $G_{\rm RP}$ are required to be smaller than 0.015\,mag.
Here, we only check the behavior of the $uv$ bands, since the photometric systematics in the other four bands are relatively small, and difficult to examine with the WD locus.
The photometric uncertainties of the $uv$ bands from SMSS DR2 are required to be smaller than 0.05\,mag.
In total, over 3000 and 4000 WDs with good photometric quality in the $u$- and $v$-band remain.

The stellar locus, $(u/v - G_{\rm RP})_0$ versus $(G_{\rm BP} - G_{\rm RP})_0$, of the selected WDs are shown in Fig.\,11.
Here, the reddening corrections are done with the SFD map.
From inspection, the WD locus from the SMSS DR2 photometry with SFD\,$E (B - V) > 0.07$ significantly deviates from that with SFD\,$E (B - V) < 0.05$, for both the $u$- and $v$-bands; these deviations are roughly consistent with the values presented in Table\,3.
In contrast, the WD loci for different SFD\,$E (B - V)$ bins given by our re-calibrated SMSS DR2 photometry are quite consistent with each other, demonstrating the power of our revised SCR method for calibrating the photometric zero-points.

Finally, Fig.\,12 shows the WD loci in the $u$-band for different $\delta$ bins. Again, the loci derived from  the SMSS DR2 photometry exhibits significant deviations for different $\delta$ bins, while the loci derived from our re-calibrated photometry agree well with each other.

\begin{table}
\centering
\caption{Fit Coefficients for the Zero-Point Magnitude Offsets as a Function of  SFD\,$E (B - V)$.}
\begin{tabular}{c|c|c}
\hline
%$\Delta X$& $a_0$ & $a_1$ & $a_2$ & $a_3$ & $a_4$ & $a_5$ & $a_6$&$a_7$\\
Coefficient & $\Delta u^{a}$ & $\Delta v^{a}$\\
&(mag)&(mag)\\
\hline
$a_0$&$-5.4720\times10^{-3}$&$2.2383\times10^{-3}$\\
$a_1$&$1.1420\times10^{-1}$&$ -5.5176\times10^{-3}$\\
$a_2$&$1.1407\times10^{1}$&$4.6240\times10^{-1}$\\
$a_3$&$-7.0705\times10^{1}$&$2.0135$\\
$a_4$&$1.8485 \times10^2$&$-1.3082\times10^1$\\
$a_5$&$ -2.4171 \times 10^2$&$2.5837\times10^1$\\
$a_6$&$1.5561 \times 10^2$&$-2.1338\times10^1$\\
$a_7$&$-3.9315 \times 10^1$&$6.3133$\\
\hline
\end{tabular}
\tablecomments {The offsets should be added to the official SMSS DR2 values.\\
$^{a} \Delta u/v = a_0 + a_1\times x + a_2\times x^2 + a_3\times x^3 + a_4\times x^4 + a_5\times x^5 + a_6\times x^6 + + a_7\times x^7$.}
\end{table}

\begin{figure}
\begin{center}
\includegraphics[scale=0.325,angle=0]{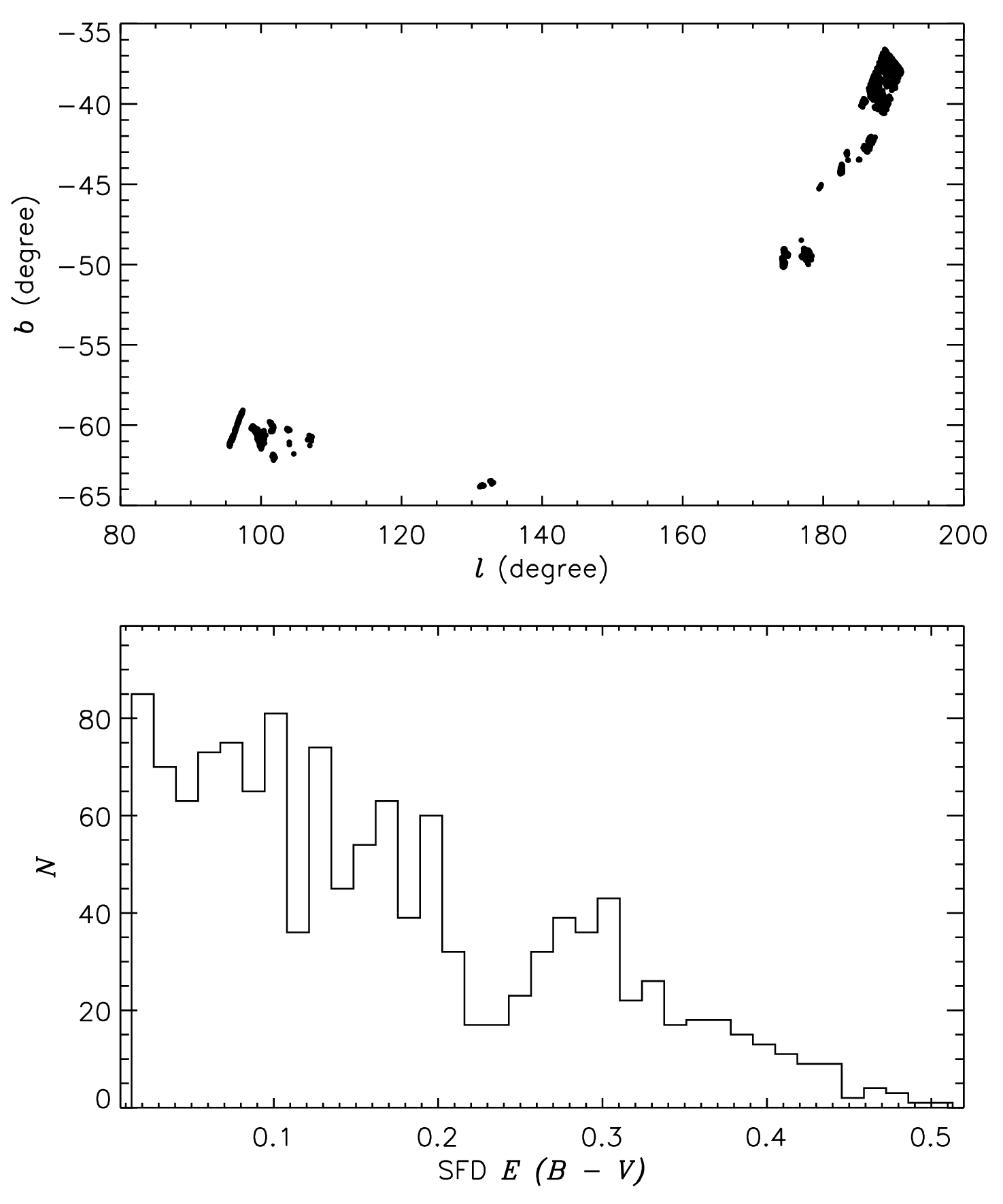}
\caption{{\it Upper panel}: Distribution of the 1291 SDSS Stripe 82 standard stars in the Galactic coordinate system. {\it Lower panel}: Number distribution of the 1291 SDSS Stripe 82 standard stars, as a function of SFD $E (B - V)$.}
\end{center}
\end{figure}

\subsection{External Checks by Comparison with SDSS Standard-Star Catalog for Stripe 82}
In this subsection, we perform another independent check of the re-calibration of the $vgriz$ bands by the SCR.
To do so, we compare the SMSS DR2 photometry to that of the SDSS standard-star catalog for Stripe 82 (Ivezi{\'c} et al. 2007).
This standard-star catalog includes $\sim$ one million candidate standard stars in the SDSS Stripe 82 region (i.e., $|\delta| \leq 1.266^{\circ}$ and $\alpha$ in the range $20^{\rm h}34^{\rm m}$--$4^{\rm h}00^{\rm m}$) with SDSS $r$-band magnitudes between 14 and 22.
The claimed overall photometric errors (including both internal zero-point errors and random photometric errors) are smaller than 0.01\,mag for stars brighter than 19.5, 20.5, 20.5, 20, and 18.5 in SDSS $ugriz$, respectively.

The SMSS DR2 catalog is cross-matched with the SDSS Stripe 82 standard-star catalog, and over 150,000 stars in common are found.
Here, we do not require all the stars in common for comparison, since most of them have low-extinction values.
We therefore take a sub-sample of 1291 stars evenly distributed over SFD $E (B - V)$ (see Fig.\,12), and with photometric errors smaller than 0.02\,mag in the SMSS $griz$ bands.
To examine the SMSS DR2 photometric zero-point accuracy, we perform the band-to-band fit as follows:
\begin{equation}
    X^{\rm SM}_0 = a_0 + a_1X^{\rm SDSS}_0 + a_2(g - i)^{\rm SDSS}_0\text{,} 
\end{equation}
where $X^{\rm SM}$ represents $v$/$g$/$r$-band and $X^{\rm SDSS}$ represents $u$/$g$/$r$-band, correspondingly.
The extinction corrections are done with the SFD map, since all of the stars in common have $b < -20^{\circ}$.
The extinction coefficients for the SDSS filters are adopted from Yuan et al. (2013).
The standard deviation of the fitting residual, $\sigma_{\rm fit}$, has contributions from the intrinsic scatter, $\sigma_{\rm int}$, the extinction-correction errors, $\sigma_{\rm ext}$, the photometric random errors, $\sigma_{\rm SM}^{\rm ph}$ and $\sigma_{\rm SDSS}^{\rm ph}$, and the zero-point errors, $\sigma_{\rm SM}^{\rm ZP}$ and $\sigma_{\rm SDSS}^{\rm ZP}$.
For SMSS DR2 and our re-calibration version, the only difference that could affect the $\sigma_{\rm fit}$ is the zero-point error of SMSS $\sigma_{\rm SM}^{\rm ZP}$.
Given the accurate photometry of SDSS Stripe 82 standard stars, if our re-calibration indeed reduces the SMSS zero-point errors, then the $\sigma_{\rm fit}$ achieved by the re-calibrated photometry should be smaller than that from the original photometry. 

The fit results are shown in Table\,6.
The values of $\sigma_{\rm fit}^{\rm Re}$ from the re-calibrated photometry are indeed all smaller than those of $\sigma_{\rm fit}^{\rm Org}$ for the original SMSS DR2.
The improvement of the re-calibrated photometry for the $v$ band is largely from the removal of the significant redenning-dependent zero-point offsets.
For the $g$-band, the reduction of $\sigma_{\rm fit}$ is not significant.
This is because the zero-point offset of the original $g$-band photometry is very small, given the SFD $E (B -V)$ range and spatial coverage of the examined sample (see Fig.\,8).
The reduction of $\sigma_{\rm fit}$ for the $r$-band in the re-calibrated photometry is mainly due to the proper correction of the spatial zero-point offset (see Fig.\,8). 

\begin{table*}
\centering
\caption{Fit Coefficients for the Zero-Point Magnitude Differences as a Function of  $\delta$ for the $uv$ Bands and $b$ for the $gr$ Bands}
\begin{tabular}{c|c|c|c|c|c|c|c}
\hline
\hline
$\Delta X$& $a_0$ & $a_1$ & $a_2$ & $a_3$ & $a_4$ & $a_5$ & Note\\
(mag)&&&&&&&\\
\hline
\multirow{3}{*}{$u$}&$-5.96\times10^{-3}$&$1.11\times10^{-3}$&$-8.00\times10^{-6}$&$-1.31\times10^{-6}$&$-2.21\times10^{-8}$&$-1.31\times10^{-10}$&$\alpha > 300^{\circ}$\\
&$1.74\times10^{-3}$&$4.52\times10^{-4}$&$-2.47\times10^{-5}$&$-2.04\times10^{-6}$&$-3.83\times10^{-8}$&$-2.33\times10^{-10}$&$\alpha < 180^{\circ}$ or $260^{\circ} < \alpha < 300^{\circ}$\\
&$-2.51\times10^{-2}$&$2.49\times10^{-3}$&$1.82\times10^{-4}$&$4.23\times10^{-6}$&$4.82\times10^{-8}$&$2.13\times10^{-10}$&$180^{\circ} < \alpha < 260^{\circ}$\\
\hline
\multirow{3}{*}{$v$}&$-1.45\times10^{-2}$&$-9.99\times10^{-4}$&$-8.44\times10^{-5}$&$-2.13\times10^{-6}$&$-1.58\times10^{-8}$&--&$\alpha > 300^{\circ}$\\
&$-1.67\times10^{-3}$&$-2.00\times10^{-4}$&$6.18\times10^{-6}$&$1.82\times10^{-7}$&$6.56\times10^{-10}$&--&$\alpha < 180^{\circ}$ or $260^{\circ} < \alpha < 300^{\circ}$\\
&$-1.76\times10^{-2}$&$1.67\times10^{-3}$&$1.11\times10^{-4}$&$1.61\times10^{-6}$&$6.36\times10^{-9}$&--&$180^{\circ} < \alpha < 260^{\circ}$\\
\hline
\multirow{4}{*}{$g$}&$-1.25\times10^{-3}$&$-1.13\times10^{-2}$&$-5.38\times10^{-1}$&$4.26\times10^{-5}$&--&--&($180^{\circ} \leq l < 330^{\circ}$) and $b > 0^{\circ}$\\
&$2.64\times10^1$&$-2.64\times10^{-1}$&$-1.78\times10^{-4}$&$-6.01\times10^{-5}$&--&--&($l \leq 180^{\circ}$ or $l > 330^{\circ}$) and $b > 0^{\circ}$\\
&$5.06\times10^{-3}$&$-2.57\times10^{-2}$&$-6.93\times10^{-1}$&$5.54\times10^{-5}$&--&--&($180^{\circ} \leq l < 330^{\circ}$) and $b < 0^{\circ}$\\
&$1.12\times10^{1}$&$-1.12\times10^{1}$&$-3.75\times10^{-4}$&$-1.61\times10^{-5}$&--&--&($l \leq 180^{\circ}$ or $l > 330^{\circ}$) and $b < 0^{\circ}$\\
\hline
\multirow{4}{*}{$r$}&$2.68$&$-2.69$&$-1.04\times10^{-3}$&$3.07\times10^{-5}$&--&--&($180^{\circ} \leq l < 330^{\circ}$) and $b > 0^{\circ}$\\
&$4.30\times10^{1}$&$-4.30\times10^{1}$&$-1.78\times10^{-4}$&$-1.26\times10^{-4}$&--&--&($l \leq 180^{\circ}$ or $l > 330^{\circ}$) and $b > 0^{\circ}$\\
&$6.43\times10^{-4}$&$-3.37\times10^{-2}$&$-1.32$&$-2.11\times10^{-5}$&--&--&($180^{\circ} \leq l < 330^{\circ}$) and $b < 0^{\circ}$\\
&$4.23\times10^{1}$&$-4.24\times10^{1}$&$-1.70\times10^{-4}$&$1.58\times10^{-5}$&--&--&($l \leq 180^{\circ}$ or $l > 330^{\circ}$) and $b < 0^{\circ}$\\
\hline
\hline
\end{tabular}
\tablecomments {The offsets should be added to the official SMSS DR2 values.}
\end{table*}

\begin{table}
\centering
\caption{Standard Deviations of the Fit Residuals}
\begin{tabular}{cccc}
\hline
$X^{\rm SM}$&$v$&$g$&$r$\\
&(mag)&(mag)&(mag)\\
\hline
$\sigma_{\rm fit}^{\rm Org}$&0.0662&0.0181&0.0175\\
$\sigma_{\rm fit}^{\rm Re}$&0.0558&0.0180&0.0167\\
\hline
\end{tabular}
\end{table}

\section{Discussion and Conclusions}

\subsection{Caveats in the Current Re-calibration}

By employing a revised SCR technique, $Gaia$ DR2 photometry, and GALAH+ DR3 stellar parameters, we have re-calibrated the SMSS DR2 photometry from O19, and achieve a zero-point accuracy of $< 1$\% for most of the comparison fields.
However, there remain several caveats that should be mentioned.   

First, the current re-calibration has only been performed for the present GALAH+ DR3 footprint, not for the full sky coverage of the SMSS DR2 photometry.
Some unknown photometric systematics could be  present in the regions not covered by GALAH+ DR3, and thus still remain in the re-calibrated photometry. 
To solve this problem, we could implement the uber-calibration approach in order to achieve a homogenous internal calibration for the whole SMSS DR2 sky coverage by using the overlapping regions.  
In addition, future GALAH observations could help with the re-calibration of the missing regions of the SMSS DR2 catalog.

Secondly, the current revised SCR method assumes constant reddening coefficients for the $Gaia$ and SkyMapper passbands for stars in different locations (environments) and spectral energy distributions (SEDs); the latter is particularly true for the $Gaia$ broad-band photometry. 
Our current assumptions for the reddening coefficients obviously could contribute some systematics in the re-calibration process.
However, we emphasize that this error should be small compared to the existing systematics in the SMSS DR2 photometry from O19.
Most of the re-calibrated fields belong to ``normal" diffuse environments, and they should follow a universal extinction law with a small scatter (e.g., Schlafly et al. 2010; Yuan et al. 2013).
In addition, most of the dwarfs we adopted to perform the re-calibration are F/G-type stars having a narrow color range in $(G_{\rm BP} - G_{\rm RP})_0$ (mainly between 0.6 and 0.8;  see Fig.\,3).
The extinction-coefficient variations due to the various stellar SEDs are therefore expected to be quite small.
Another critical issue is the significant discrepancies of the extinction coefficients in some bands (i.e., $iz$-bands here; see Table\,2) predicted by different extinction laws.
Our SCR technique cannot present a conclusive re-calibration for the $iz$ photometry in SMSS DR2 due to this issue.
As a next step, those issues should be properly considered and solved, to further refine the photometric zero-point accuracy of the SMSS photometry.

%Finally, we do not presently anchor the scale of the re-calibrated photometry to the well-defined AB or Vega photometric systems.
%As mentioned previously, this could be accomplished by comparing the re-calibrated photometry to a few well-defined photometric standard stars.
%However, there are as yet no well-defined photometric standard stars for the SkyMapper filter system.
%Another approach would be to integrate synthetic SkyMapper magnitudes on the AB/Vega photometric systems using the spectro-photometric standard stars from the NGSL and the CALSPEC spectral library (Bohlin, Gordon \& Tremblay 2014).
%However, as mentioned previously, most of those stars are too bright, and saturate in the SMSS DR2 photometry (especially in the $griz$ bands).

\subsection{Conclusions and Perspectives on SCR+Gaia for Future Large-Scale Photometric Surveys}
We have applied a revised SCR technique, together with $Gaia$ DR2 photometry and  GALAH+ DR3 stellar atmospheric parameters, to re-calibrate the zero-points of the SMSS DR2 photometry, reported by O19.
As we expected, strong reddening-dependent zero-point systematics are found for all the SkyMapper $uv$ bands.
The photometric zero-points are close to zero in low-extinction regions, and then gradually increase with increasing SFD $E (B - V)$, and can reach as large as 0.174 and 0.134\,mag for $uv$, respectively, in high-extinction regions with SFD\,$E (B - V) \sim 0.50$.
This reddening-dependent trend of the photometric zero-points is largely caused by the dust terms in the passband transformations used to construct the photometric calibrators used by O19 for SMSS DR2.
For the $gr$ bands, the zero-points show negligible variations with SFD\,$E(B - V)$, given their tiny coefficients on the dust term in the transformation.
Our study also demonstrates the existence of small, but significant, spatial variations of the zero-points for all the $uvgr$ bands. 
By properly correcting for the reddening-dependent and spatial zero-point systematics, most of the calibrated fields have zero-point uncertainties smaller than 10\,mmag. Independent checks using Str{\"o}mgren photometry, WD loci and the SDSS Stripe 82 standard-star catalog also show the power of our revised SCR method for calibrating the SMSS DR2 photometry.

In the current work, we have demonstrated the advantages of the revised SCR method for calibrating modern digital photometric surveys to $< 1$\% accuracy.
The basis of this technique relies on massive large-scale spectroscopic surveys used in conjunction with an independent all-sky uniform photometric survey (in this case $Gaia$ DR2, with expected improved results in the near future).
In the Northern Hemisphere, the LAMOST spectroscopic surveys have obtained over ten million low-resolution ($R \sim 1800$) spectra covering most of the Northern sky ($-10 < \delta < 60^{\circ}$).  We emphasize, however, that improvements in the determination of stellar atmospheric parameters, in particular [Fe/H], for very low-metallicity stars ([Fe/H] $< -2.0$) for LAMOST stars will be very useful (see, e.g., Yuan et al. 2020, for refined parameters for LAMOST DR3 VMP stars; similar improvements are being obtained for LAMOST DR5 VMP stars at present).
In the Southern Hemisphere, we have the ongoing GALAH spectroscopic survey and the 4MOST (de Jong et al. 2019) survey expected to begin in the relatively  near future.

One important limitation of the current and future photometric surveys is that they have too-faint bright limits, compared to the bright limits of many massive spectroscopic surveys.
We therefore encourage the ongoing/planned photometric surveys to execute short-exposure, shallow surveys first (or in tandem, as is being done with the S-PLUS Ultra Short Survey; see  Mendes de Oliveira et al. 2019). After calibrating those shallow surveys with our revised SCR technique, they become the second-level photometric standards for the fainter main surveys.

To conclude, we believe the revised SCR approach will be a promising method to calibrate many of the ongoing/planned large-scale digital sky surveys (e.g., LSST, J-PAS, J-PLUS, S-PLUS, SAGE, and Mephisto) to achieve photometric zero-points at the few mmag level. As we have emphasized, the most important photometric bands for estimation of metallicity, the $uv$ bands in SMSS DR2 and the Ca~II HK filters in other surveys (e.g., Pristine; Starkenburg et al. 2017, and J-PLUS/ S-PLUS) are very sensitive to zero-point offsets.  If properly calibrated, these photometric surveys will enable determination of [Fe/H] estimates down to as low as [Fe/H] $\sim -3.5$, making it possible for high-resolution spectroscopic follow-up efforts to concentrate on the most metal-poor stars ([Fe/H] $< -2.0$), which contain precious information on the nucleosynthesis products of the very first generations of stars (e.g., Howes et al. 2015; Nordlander et al. 2019).

Finally, our re-calibrations can also be applied to SMSS DR3, which adopted the same photometric calibration strategy as in DR2 (O19).
We also provide the routines and scripts on GitHub  (\url{github.com/comhy/SMSS-DR2-ZP-corrections}) for the corrections of the zero-points of SMSS DR2/3 photometry.

 \section*{Acknowledgements} 
 This work is supported by National Natural Science Foundation of China grants 11903027,  11833006, and U1731108, and  National Key R\&D Program of China No. 2019YFA0405503. 
Y.H. is supported by the Yunnan University grant C176220100006. 
C.A.O. acknowledges support from the Australian Research Council through Discovery Project DP190100252.
T.C.B. acknowledges partial support from grant PHY 14-30152, Physics
Frontier Center/JINA Center for the Evolution of the
Elements (JINA-CEE), awarded by the US National Science
Foundation. His participation in this work was initiated by conversations that took place during a visit to China in 2019, supported by a PIFI Distinguished Scientist award from the Chinese Academy of Science.
L.C. acknowledges support from the Australian Research Council grants FT160100402.
A. D. M. is supported by an ARC Future Fellowship (FT160100206).
Y.S.T. is supported  by the NASA Hubble Fellowship grant HST-HF2-51425.001 awarded by the Space Telescope Science Institute.

We acknowledge the use of data reduction infrastructure developed within the GALAH Survey collaboration. 
The GALAH survey web site is www.galah-survey.org. 
This work is based on data acquired through the Australian Astronomical Observatory, under programmes: A/2014A/25, A/2015A/19, A2017A/18 (The GALAH survey); A/2015A/03, A/2015B/19, A/2016A/22, A/2016B/12, A/2017A/14 (The K2-HERMES K2-follow-up program);
A/2016B/10 (The HERMES-TESS program); A/2015B/01 (Accurate physical parameters of Kepler K2 planet search targets); S/2015A/012 (Planets in clusters with K2). 

The national facility capability for SkyMapper has been funded through ARC LIEF grant LE130100104 from the Australian Research Council, awarded to the University of Sydney, the Australian National University, Swinburne University of Technology, the University of Queensland, the University of Western Australia, the University of Melbourne, Curtin University of Technology, Monash University and the Australian Astronomical Observatory. SkyMapper is owned and operated by The Australian National University's Research School of Astronomy and Astrophysics. The survey data were processed and provided by the SkyMapper Team at ANU. The SkyMapper node of the All-Sky Virtual Observatory (ASVO) is hosted at the National Computational Infrastructure (NCI). Development and support the SkyMapper node of the ASVO has been funded in part by Astronomy Australia Limited (AAL) and the Australian  Government through the Commonwealth's Education Investment Fund (EIF) and National Collaborative Research Infrastructure Strategy (NCRIS), particularly the National eResearch Collaboration Tools and Resources (NeCTAR) and the Australian National Data Service Projects (ANDS).

Parts of this research were supported by the Australian Research Council Centre of Excellence for All Sky Astrophysics in 3 Dimensions (ASTRO 3D), through project number CE170100013.

This work has made use of data from the European Space Agency (ESA) mission
{\it Gaia} (https://www.cosmos.esa.int/gaia), processed by the {\it Gaia}
Data Processing and Analysis Consortium (DPAC,
https://www.cosmos.esa.int/web/gaia/dpac/consortium). Funding for the DPAC
has been provided by national institutions, in particular the institutions
participating in the {\it Gaia} Multilateral Agreement.

\vfill\eject

\vfill\eject

%\appendix
%
%\section{Tables for Correcting the Reddening-Dependent and Spatial Zero-Point Errors in the SMSS DR2 Photometry}
%
%Table\,A1 lists the magnitude zero-point offsets, as a function of SFD\,$E (B - V)$.
%Table\,A2 lists the fit coefficients for the magnitude zero-point offsets, as a function of $\delta$ and $b$, for the $uv$ and $griz$ bands, respectively.

\end{document}